\documentclass[twoside,twocolumn,9pt]{article}
\usepackage[super,sort&compress,comma]{natbib} 
\usepackage[version=3]{mhchem}
\usepackage[left=1.5cm, right=1.5cm, top=1.785cm, bottom=2.0cm]{geometry}
\usepackage{balance}
\usepackage{mathptmx}
\usepackage{sectsty}
\usepackage{graphicx} 
\usepackage{lastpage}
\usepackage[format=plain,justification=justified,singlelinecheck=false,font={stretch=1.125,small,sf},labelfont=bf,labelsep=space]{caption}
\usepackage{float}
\usepackage{fancyhdr}
\usepackage{fnpos}
\usepackage[english]{babel}
\usepackage{array}
\usepackage{droidsans}
\usepackage{charter}
\usepackage[T1]{fontenc}
\usepackage[dvipsnames]{xcolor}
\usepackage{setspace}
\usepackage[compact]{titlesec}
\usepackage[hyperfootnotes=false]{hyperref}

\hypersetup{
     colorlinks   = true,
     citecolor    = blue
}

\usepackage[T1]{fontenc}
\usepackage{lmodern}

\usepackage{epstopdf}

\definecolor{cream}{RGB}{222,217,201}

\begin{document}

\pagestyle{fancy}
\thispagestyle{plain}
\fancypagestyle{plain}{
\renewcommand{\headrulewidth}{0pt}
}

\makeFNbottom
\makeatletter
\renewcommand\LARGE{\@setfontsize\LARGE{15pt}{17}}
\renewcommand\Large{\@setfontsize\Large{12pt}{14}}
\renewcommand\large{\@setfontsize\large{10pt}{12}}
\renewcommand\footnotesize{\@setfontsize\footnotesize{7pt}{10}}
\makeatother

\renewcommand{\thefootnote}{\fnsymbol{footnote}}
\renewcommand\footnoterule{\vspace*{1pt}%
\color{cream}\hrule width 3.5in height 0.4pt \color{black}\vspace*{5pt}} 
\setcounter{secnumdepth}{5}

\makeatletter 
\renewcommand\@biblabel[1]{#1}            
\renewcommand\@makefntext[1]%
{\noindent\makebox[0pt][r]{\@thefnmark\,}#1}
\makeatother 
\renewcommand{\figurename}{\small{Fig.}~}
\sectionfont{\sffamily\Large}
\subsectionfont{\normalsize}
\subsubsectionfont{\bf}
\setstretch{1.125} 
\setlength{\skip\footins}{0.8cm}
\setlength{\footnotesep}{0.25cm}
\setlength{\jot}{10pt}
\titlespacing*{\section}{0pt}{4pt}{4pt}
\titlespacing*{\subsection}{0pt}{15pt}{1pt}

\fancyfoot{}
\fancyfoot[LO,RE]{\vspace{-7.1pt}\includegraphics[height=9pt]{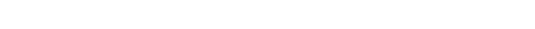}}
\fancyfoot[CO]{\vspace{-7.1pt}\hspace{11.9cm}\includegraphics{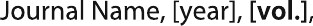}}
\fancyfoot[CE]{\vspace{-7.2pt}\hspace{-13.2cm}\includegraphics{head_foot/RF}}
\fancyfoot[RO]{\footnotesize{\sffamily{1--\pageref{LastPage} ~\textbar  \hspace{2pt}\thepage}}}
\fancyfoot[LE]{\footnotesize{\sffamily{\thepage~\textbar\hspace{4.65cm} 1--\pageref{LastPage}}}}
\fancyhead{}
\renewcommand{\headrulewidth}{0pt} 
\renewcommand{\footrulewidth}{0pt}
\setlength{\arrayrulewidth}{1pt}
\setlength{\columnsep}{6.5mm}
\setlength\bibsep{1pt}

\makeatletter 
\newlength{\figrulesep} 
\setlength{\figrulesep}{0.5\textfloatsep} 
\newcommand{\mos}{MoS$_2$}
\newcommand{\mo}{moir\'{e}}
\newcommand{\Mo}{Moir\'{e}}
\def\angstrom{Å}
\newcommand{\topfigrule}{\vspace*{-1pt}%
\noindent{\color{cream}\rule[-\figrulesep]{\columnwidth}{1.5pt}} }

\newcommand{\botfigrule}{\vspace*{-2pt}%
\noindent{\color{cream}\rule[\figrulesep]{\columnwidth}{1.5pt}} }

\newcommand{\dblfigrule}{\vspace*{-1pt}%
\noindent{\color{cream}\rule[-\figrulesep]{\textwidth}{1.5pt}} }

\makeatother

\twocolumn[
  \begin{@twocolumnfalse}
{\includegraphics[height=30pt]{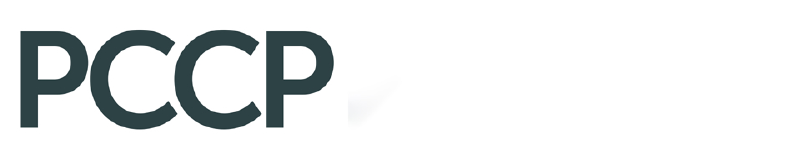}\hfill\raisebox{0pt}[0pt][0pt]{\includegraphics[height=55pt]{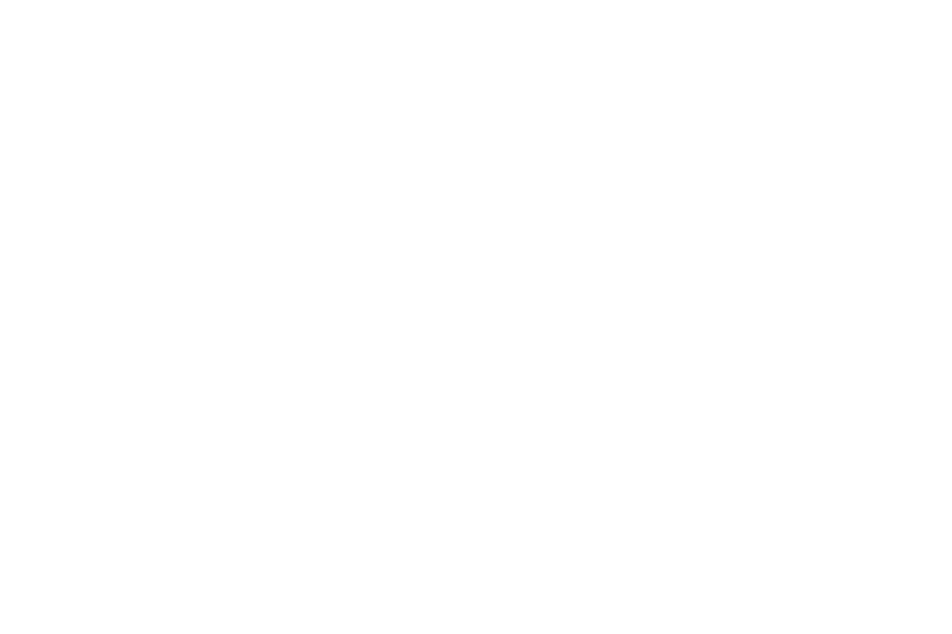}}\\[1ex]
\includegraphics[width=18.5cm]{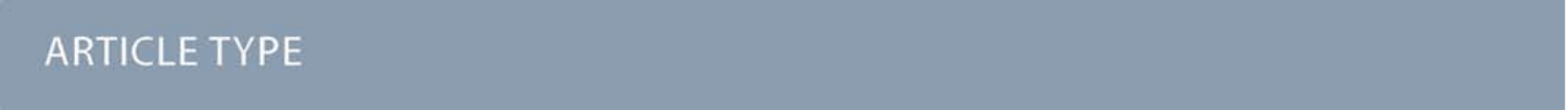}}\par
\vspace{1em}
\sffamily
\begin{tabular}{m{4.5cm} p{13.5cm} }

\includegraphics{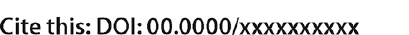} & \noindent\LARGE{\textbf{Review of the tight-binding method applicable to the properties of moir\'e superlattices$^\dag$}} \\
\vspace{0.3cm} & \vspace{0.3cm} \\

 & \noindent\large{Xueheng Kuang,\textit{$^{a}$} Federico Escudero,\textit{$^{b}$} Pierre A. Pantale\'on,\textit{$^{b}$} Francisco Guinea\textit{$^{bc}$} and Zhen Zhan$^{\ast}$\textit{$^{b}$}} \\

\includegraphics{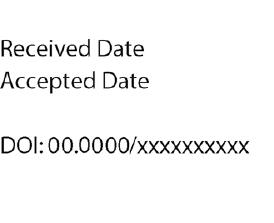} & \noindent\normalsize{Moir\'{e} superlattices have emerged as a versatile platform for exploring a wide range of exotic quantum phenomena. Unlike angstrom-scale materials, the moir\'{e} length-scale system contains a large number of atoms, and its electronic structure is significantly modulated by the lattice relaxation. These features pose a huge theoretical challenge. Among the available theoretical approaches, tight-binding (TB) methods are widely employed to predict the electronic, transport, and optical properties of systems such as twisted graphene, twisted transition-metal dichalcogenides (TMDs), and related moiré materials. In this review, we provide a comprehensive overview of atomistic TB Hamiltonians and the numerical techniques commonly used to model graphene-based, TMD-based and hBN-based moir\'{e} superlattices. We also discuss the connection between atomistic TB descriptions and effective low-energy continuum models. Two examples of different moir\'{e} materials and geometries are provided to emphasize the advantages of the TB methods. This review is intended to serve as a theoretical and practical guide for those seeking to apply TB methods to the study of various properties of moiré superlattices.} \\

\end{tabular}

 \end{@twocolumnfalse} \vspace{0.6cm}

  ]

\renewcommand*\rmdefault{bch}\normalfont\upshape
\rmfamily
\section*{}
\vspace{-1cm}


\footnotetext{\textit{$^{a}$~Texas Materials Institute and Department of Mechanical Engineering, The University of Texas at Austin, Austin, Texas 78731, United States. }}
\footnotetext{\textit{$^{b}$~IMDEA Nanociencia, Faraday 9, Madrid 28015, Spain. E-mail: zhenzhanh@gmail.com}}
\footnotetext{\textit{$^{c}$~Donostia International Physics Center, Paseo Manuel de Lardizabal 4, San Sebastian 20018, Spain.}}



\section{Introduction}

Moiré superlattices can be constructed by stacking two-dimensional materials with relative rotation or slight lattice mismatch, giving rise to long-wavelength interference patterns in their atomic structures\cite{andrei2021marvels,he2021moire}. A famous example is twisted bilayer graphene (TBG), where a small-angle rotation between the graphene layers generates a moiré superlattices with emergent electronic properties, for example, moiré flat band structure\cite{andrei2020graphene,cai2021fabrication,carr2020electronic}. Such moiré materials have rapidly become a versatile platform for exploring exotic physics\cite{andrei2021marvels,behura2021moire}, as well as new opportunities in materials science and chemistry \cite{wang2023moire,yu2022tunable,hsieh2023domain,yu2024twisted}. Remarkably, experiments have revealed a variety of strongly correlated phenomena and topology in these systems. Unconventional superconductivity and correlated insulating states have been observed in twisted bilayer graphene layers\cite{cao2018unconventional,cao2018correlated,yankowitz2019tuning,burg2019correlated,saito2020independent,arora2020superconductivity,nuckolls2020strongly}, multilayer graphene/hexagonal boron nitride (hBN) heterostructures\cite{chen2019evidence,sun2021correlated,yang2022spectroscopy}, as well as in moiré transition-metal dichalcogenides (TMDs)\cite{xu2020correlated,wang2020correlated,guo2025superconductivity,xia2025superconductivity}. Beyond superconductivity, moiré systems exhibit tunable ferromagnetism\cite{chen2020tunable,tschirhart2021imaging,song2021direct,lin2022spin,wang2022light}, ferroelectricity\cite{zheng2020unconventional,zhang2023visualizing,ding2024engineering}, and integer and fractional quantum anomalous Hall effects \cite{serlin2020intrinsic,li2021quantum,park2023observation,cai2023signatures,xu2023observation,lu2024fractional,ju2024fractional,lu2025extended}. These experimental breakthroughs highlight the potential of moiré materials for applications in quantum technologies and optoelectronics\cite{nowakowski2025single,du2025ultrasensitive}, including quantum computing\cite{jha2025large,song2025twisted,zheng2024gate,kezilebieke2022moire,devries2021gate}, lasing and cavity engineering \cite{qian2024lasing,zhang2024moire,yang2024shg,wang2025cavity,nowakowski2025single}, and chemical property tuning via twist-angle control \cite{yu2022tunable,yu2022tuning,wang2023moire,schleder2023one,zhan2025moire}.

Experimental observations on moiré materials also motivate extensive theoretical and numerical efforts to understand these phenomena and provide accurate and robust predictions of the moiré systems. However, theoretical modeling remains challenging because realistic moir\'{e} superlattices often contain thousands of atoms\cite{carr2020electronic}. In addition, lattice reconstruction and atomic relaxation play critical roles in determining electronic, transport and optical properties of moiré materials\cite{rosenberger2020twist,li2021lattice,li2021imaging,tilak2022moire,halbertal2022unconventional,zhao2023excitons,van2023rotational,fu2024lattice}. Several atomistic approaches have been employed to study the electronic structure of moir\'{e} superlattices. Density functional theory (DFT) not only supports phenomenological descriptions and synthesis control across diverse two-dimensional (2D) materials\cite{sfuncia2023graphiticgan,filho2024inaln,mounet2018exfoliation}, but also provides accurate descriptions of their electronic structures\cite{haastrup2018c2db}, and has been applied to relatively small and large-angle twised graphene layers, twisted bilayer TMDs and twisted bilayer hBN\cite{haddadi2020moire,zhao2020formation,naik2018ultraflatbands,venkateswarlu2020electronic,devakul2021magic,zhang2021electronic,kundu2022moire,jia2024moire,xu2025multiple}. However, its computational cost makes the direct simulation of large-scale moiré superlattices inefficient. 
At the opposite limit, continuum models offer effective low-energy descriptions that capture essential band features and have been widely used to provide insights into some experimental observations\cite{bistritzer2011moire,lopes2012continuum,chang2023continuum,mao2024transfer,morales2024magic}. 

Bridging these two methods, the tight-binding (TB) model offers an atomistic yet computationally efficient framework for modeling moiré materials\cite{foulkes1989tight,bowler1997comparison,papaconstantopoulos2003slater,cui2014density}. Crucially, atomic TB Hamiltonians have been built to simulate a broad variety of 2D materials such as graphene\cite{mccann2013electronic}, TMDs\cite{cappelluti2013mos2,liu2013threeband}, black phosphorus\cite{rudenko2014phosphorene}, and group-IV/V “enes” (silicene/germanene/stanene)\cite{nakhaee2019single,chegel2020tunable,rahman2020electronic}. Unlike continuum models, TB retains lattice-level resolution, making it possible to capture the effects of atomic relaxation\cite{leconte2022relaxation,kuang2021collective}, local disorder\cite{shi2020large,ren2023real}, strain\cite{mannai2021twistronics}, and chemical specificity\cite{slater1954simplified,mehl1996nrltb,jancu1998sp3d5s}. Furthermore, the method can be systematically extended to include many-body interactions\cite{pons2020flat,gonzalez2020time,bhowmik2023spin}, external fields\cite{wu2021lattice}, and coupling to lattice or optical degrees of freedom\cite{wang2022polarization}. Moreover, TB model is orders of magnitude more efficient than DFT method, enabling conventional numerical simulations of realistic moiré supercells with thousands of atoms. The TB model can be further integrated with advanced real-space linear scaling numerical techniques to simulate up to millions of atoms\cite{bowler1997comparison,yuan2010modeling,li2023tbplas,joao2020kite}. Because of this unique balance between accuracy and efficiency, TB method has become a central tool for studying electronic, transport, and optical properties of moiré superlattices across material platforms, from twisted bilayer graphene to hBN- and TMDs-based heterostructures.

In this Review, we focus on TB methods that have been applied to study broad properties of moiré materials such as electronic, transport, and dynamical properties. In Section 2, we  introduce widely used TB Hamiltonians in moiré materials including graphene-based, TMDs-based and hBN-based moiré superlattices. In Section 3, we also review the 
numerical methods in dealing with the large scale TB Hamiltonian matrices and introduce some practical software packages used to study the properties of moiré materials. In Section 4 we then analyze the relation of TB methods to DFT and continuum models used in moiré materials. We also display two typical examples of implementing TB methods to study properties of moir\'{e} materials in section 5.\\

\section{Tight-binding Hamiltonian of Moir\'e materials}

In the study of two-dimensional (2D) superlattices, the most commonly investigated materials are graphene, hBN, and transition metal dichalcogenides (TMDs). In homobilayer systems, such as TBG, the superlattice structure is characterized by a single twist angle $\theta$. For certain special twist angles, the superlattice preserves translational symmetry and forms a well-defined commensurate supercell. At these special angles, the two graphene lattices \textit{beat} in space, giving rise to a moiré period defined by integer numbers of graphene lattice vectors. We refer to these angles as commensurate angles. Another structure of interest is an incommensurate structure, the dodecagonal quasicrystal with $\theta = 30^{\circ}$~\cite{yao2018quasicrystalline,ahn2018dirac,li2024tuning}. The atomistic TB model is widely used to study the electronic structures of these superlattices. The starting point for the TB model is the construction of the superlattice. Therefore, in this section, we will first give a brief description of the geometry and then explicitly discuss the TB Hamiltonians of these systems. For simplicity, we limit our attention to moir\'e systems. The TB Hamiltonian for the incommensurate case is straightforward.

\subsection{Moir\'e geometry} 
 
A moir\'e pattern can be generated in several ways. For example, when two single layers of 2D materials are stacked on top of each other with a relative commensurate angle, a moir\'e pattern is formed~\cite{lopes2007graphene}. Moir\'e patterns can also be created solely by applying strain~\cite{escudero2024designing}. The period of the moir\'e pattern is determined by the twist angle or lattice mismatch. In this section, we briefly introduce the geometry of the moiré pattern defined by rotation. The general and universal formulas for generating moiré systems are given in Refs.~\cite{escudero2024designing,shi2020large}.

\begin{figure}[t]
    \centering
    \includegraphics[width=0.5\textwidth]{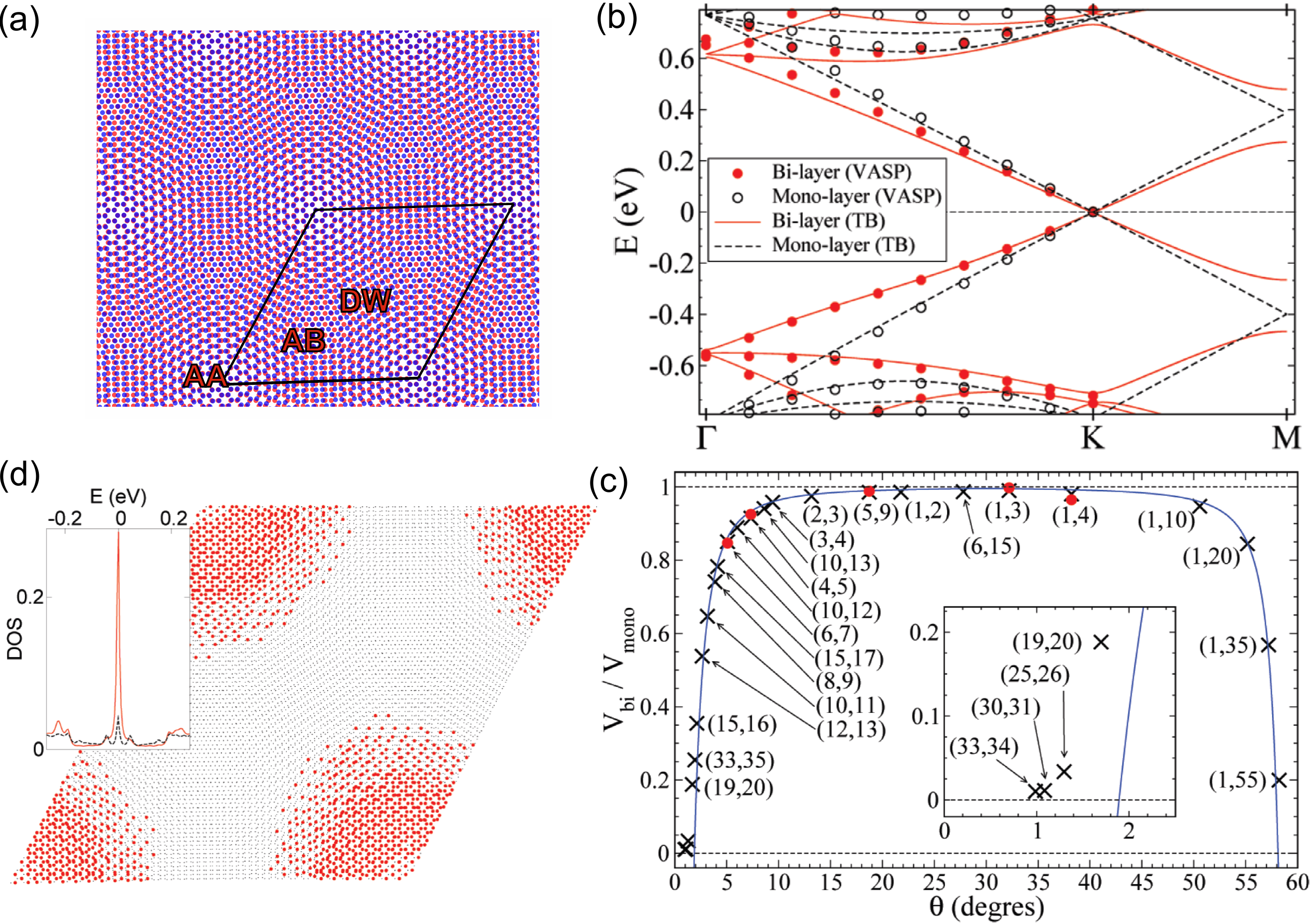}
    \caption{(a) The atomic structure of TBG with $\theta=3.15^{\circ}$. The moir\'e unit cell is illustrated with a black parallelogram. (b) Band structure of TBG with $\theta=5.08^{\circ}$ obtained by performing TB (solid line) and ab initio (dot) calculations. In the TB calculation, the hopping parameters are $t_0 = 2.7 eV$ and $t_1 = 0.48 eV$. (c) Fermi velocity ratio $V_{bi}/V_{mono}$ of TBG versus angle $\theta$. Red dot for the ab initio calculations and black cross for the TB calculations. The velocity close to 0 at angle $\theta = 1.08^\circ$ with integer pair $(30,31)$. (d) Distribution of one eigenstate at K point with energy $E=0$, in the unit cell of TBG with  $\theta = 1.08^\circ$. Black small dots are the positions of all atoms, red dots are atoms where 80\% of the states are localized. Inset shows the local density of states (DOS) of the AA stacking (solid red line) and the total DOS (dashed black line). Adapted with permission from \cite{trambly2010localization}. Copyright (2010) American Chemical Society.}
    \label{fig:S-p TB}
\end{figure}

For the TBG case, the period of the moir\'e pattern is\cite{van2015relaxation}:
\begin{equation}
A_m = \frac{a_G}{2|\sin{\theta/2}|},
\end{equation}
where $a_G$ is the graphene lattice constant. The TBG could be constructed by identifying a common periodicity in the two graphene monolayers. For one layer, we define a supercell with a lattice vector $\mathbf A_1=n \mathbf a_1 + m \mathbf a_2$, where $\mathbf a_{1,2}$ are the lattice vectors of monolayer graphene, and $m,n$ are integers with $n>m \geq 1$. For the second layer, a supercell with the same size and rotated by an angle $\theta$ can be obtained by taking a lattice vector $\mathbf A_2=-m\mathbf a_1 + (n+m) \mathbf a_2$. The moir\'e superlattice is then constructed by rotating the cell with $\mathbf A_1$ by $\theta/2$ and the cell with $\mathbf A_2$ by $-\theta/2$. Each pair of $(n,m)$ identifies a commensurate supercell with twist angle $\theta$ as:
\begin{equation}
\cos{\theta}=\frac{1}{2}\frac{n^2+4nm+m^2}{n^2+nm+m^2}.
\end{equation}
Figure \ref{fig:S-p TB}(a) shows a moir\'e pattern of TBG with $\theta=3.15^{\circ}$, which consists of AA, AB and DW stackings. These stacking configurations have distinct stacking energies, resulting in a strong lattice reconstruction of the system to achieve an equilibrium condition. The moir\'e pattern can be visualized by means of transmission electron microscopy and scanning tunneling microscopy\cite{shi2020large,yoo2019atomic}. 

\subsection{Graphene-based moir\'e materials }
\subsubsection{Single-particle TB method}

The most widely studied moir\'e materials are graphene-based heterostructures, such as twisted bilayer graphene, twisted trilayer graphene, and twisted multilayer graphene. To describe their electronic structure, single-particle tight-binding models are commonly employed. A typical example is the TB model restricted to the $p_z$ orbital, which captures the essential low-energy physics of graphene. The Hamiltonian is written as
\begin{equation}
H_0 = \sum_i \epsilon_i c_i^{\dagger} c_i^{\phantom{\dagger}} + \sum_{i \neq j} t_{ij} c_i^{\dagger} c_j^{\phantom{\dagger}}
\label{hal0},
\end{equation}
where $\epsilon_i$ is the onsite energy of the $p_z$ orbital at site $i$, and $t_{ij}$ denotes the hopping between $p_z$ orbitals at sites $i$ and $j$. The hopping amplitudes follow the Slater–Koster (SK) relation
\begin{equation}
t_{ij} = n^2 V_{pp\sigma}(r_{ij}) + (1 - n^2) V_{pp\pi}(r_{ij}),
\label{eq:tij}
\end{equation}
where $r_{ij} = |\mathbf{r}_j - \mathbf{r}_i|$ is the distance between sites $i$ and $j$, and $n$ is the direction cosine along the $\mathbf{e}_z$ axis perpendicular to the graphene plane. The SK parameters $V_{pp\pi}$ and $V_{pp\sigma}$ are given by \cite{trambly2010localization,trambly2012numerical}
\begin{equation}
V_{pp\pi}(r_{ij}) = -t_0 e^{q_{\pi} \left(1 - r_{ij}/d\right)} F_c(r_{ij}),
\label{eq:vpppi}
\end{equation}
\begin{equation}
V_{pp\sigma}(r_{ij}) = t_1 e^{q_{\sigma} \left(1 - r_{ij}/h\right)} F_c(r_{ij}),
\label{eq:vppsigma}
\end{equation}
where $d$ and $h$ are the nearest in-plane and out-of-plane carbon–carbon distances, respectively. The parameters $t_0$ and $t_1$ set the in-plane and out-of-plane hopping strengths, while $q_{\pi}$ and $q_{\sigma}$ are decay factors satisfying $\frac{q_{\sigma}}{h} = \frac{q_{\pi}}{d} = 2.218 ,\text{\AA}^{-1}$. A smooth cutoff function
\begin{equation}
F_{c}(r)=\frac{1}{1+e^{\left(r-r_{c}\right)/l_{c}}},
\end{equation}
with $l_c=0.265 \,\text{\AA}$ and cutoff distance $r_c=5.0 \,\text{\AA}$, is used to suppress long-range hopping terms. According to Eqs. \eqref{eq:vpppi} and \eqref{eq:vppsigma}, the electronic structure varies with SK hopping parameters ($t_0$ and $t_1$) and  bond lengths ($d$ and $h$). For example, by modulating slightly the SK parameters, the first magic angle can be shifted between 1.05$^\circ$ and 1.2$^\circ$ \cite{kuang2021collective}. For bilayer graphene case, the equilibrium bond length are $d=1.419\,\text{\AA}$ and $h_{AA}=3.599\,\text{\AA}$, which are reproduced by a DFT + vdW calculation\cite{gargiulo2017structural}. More information on the bound length refers to Refs. \cite{van2015relaxation,gargiulo2017structural,guinea2019continuum,leconte2022relaxation,zakharchenko2009finite}. This minimal $p_z$-orbital model provides a reliable starting point for describing the electronic structure of graphene-based moir\'e systems.  In practice, more refined models are often required to include lattice relaxation, correlation effects, or substrate-induced modifications. 

In 2010, Guy Trambly de Laissardi\`ere and co-workers derived the above TB model and predicted the electronic structure of TBG at different twist angles~\cite{trambly2010localization}. The agreement between the \textit{ab initio} and TB results was excellent (see the red dot and red line in Fig.~\ref{fig:S-p TB}(b)). From the calculated band dispersions along $\Gamma$--$K$, they extracted the velocity of the Dirac states near the $K$ point using $V_{bi}=\frac{1}{\hbar}\frac{\partial E}{\partial k}$, and compared it with the corresponding value in monolayer graphene, $V_{mono}$. As shown in Fig.~\ref{fig:S-p TB}(c), the velocity renormalization varies symmetrically around $\theta=30^{\circ}$.

Within the small-angle regime ($\theta<3^{\circ}$), the low-energy bands become flat. At the particular twist angle $\theta=1.08^{\circ}$, referred to as the first \textit{magic angle}, the velocity tends to zero. This value is very close to $\theta = 1.05^\circ$, obtained from the continuum model by Bistritzer and MacDonald~\cite{bistritzer2011moire}, and consistent with the experimentally observed magic angle near $\theta=1.1^{\circ}$~\cite{cao2018unconventional}. In the flat-band regime, the moir\'e potential induces a strong peak near the charge neutrality point in the local density of states (DOS) of the AA stacking region, where the states are mainly localized (Fig.~\ref{fig:S-p TB}(d)). This behavior was unexpected at the time, since Dirac electrons in graphene obey the so-called Klein paradox, which makes them difficult to localize with an electrostatic potential~\cite{katsnelson2006chiral}.

A similar TB model was proposed by E.~Su\'arez Morell~\textit{et al.} in 2010, who predicted the magic angle at $1.5^{\circ}$~\cite{suarez2010flat}. Their model included up to third-nearest-neighbor interlayer hoppings. The precise value of the magic angle depends strongly on the hopping parameters $t_0$ and $t_1$~\cite{kuang2021collective}, which can be tuned in realistic models to better fit DFT results~\cite{haddadi2020moire,wu2021lattice} or experimental data~\cite{kuang2021collective}. Moreover, based on the above TB framework, the existence of flat bands has also been demonstrated in twisted trilayer graphene~\cite{lopez2020electrical,ramires2021emulating,wu2021lattice,wu2021magic,goodwin2021flat}, twisted double bilayer graphene~\cite{haddadi2020moire,culchac2020flat,liang2020effect}, and twisted multilayer graphene~\cite{perrin2024electric,foo2024extended,culchac2025flat}.

The atomistic TB model offers several advantages for studying moir\'e systems. First, lattice relaxation effects can be incorporated by modifying the distance-dependent hoppings $t_{ij}$ in Eq.~(\ref{eq:tij}) according to the relaxed atomic positions \cite{nam2017lattice,guinea2019continuum,kuang2021collective,leconte2022relaxation}, which allows the model to reproduce the observed band gaps between flat and remote bands \cite{cao2016superlattice,cao2018correlated}. One option to obtain relaxed structures is through the classical simulation package LAMMPS \cite{plimpton1995fast}. For reference, libraries of lattice relaxation are available for graphene \cite{guinea2019continuum}, TMDs \cite{naik2019kolmogorov}, and hBN \cite{li2024moire} (LAMMPS potentials are presented in Table 1).

Second, substrate effects, strain, impurities, and external electric or magnetic fields can be readily implemented within the TB framework. For example, a perpendicular electric field can be introduced by adding an onsite potential term to each site, while a perpendicular magnetic field can be included through the Peierls substitution \cite{vonsovsky1989quantum}
\begin{equation}
t_{ij} \to t_{ij} \cdot \exp\left({\mathrm{i}}\frac{e}{\hbar c}\int_i^j \mathbf{A}\cdot \mathrm{d} \mathbf{l} \right)
= t_{ij} \cdot \exp\left({\mathrm{i}}\frac{2 \pi}{\Phi_0}\int_i^j \mathbf{A}\cdot \mathrm{d} \mathbf{l} \right),
\label{Peierls}
\end{equation}
where $\int_i^j \mathbf{A}\cdot\mathrm{d}\mathbf{l}$ is the line integral of the vector potential from orbital $i$ to orbital $j$, and $\Phi_0=2\pi c\hbar/e$ is the flux quantum. For a perpendicular magnetic field along $-z$, the Landau gauge $\mathbf{A}=(By, 0, 0)$ can be used. This framework enables the study of large-scale properties such as the quantum Hall effect in twisted graphene using linear-scaling methods and linear-response theory \cite{joao2020kite,li2023tbplas}.

\begin{figure}[t]
 \centering
 \includegraphics[width=0.4\textwidth]{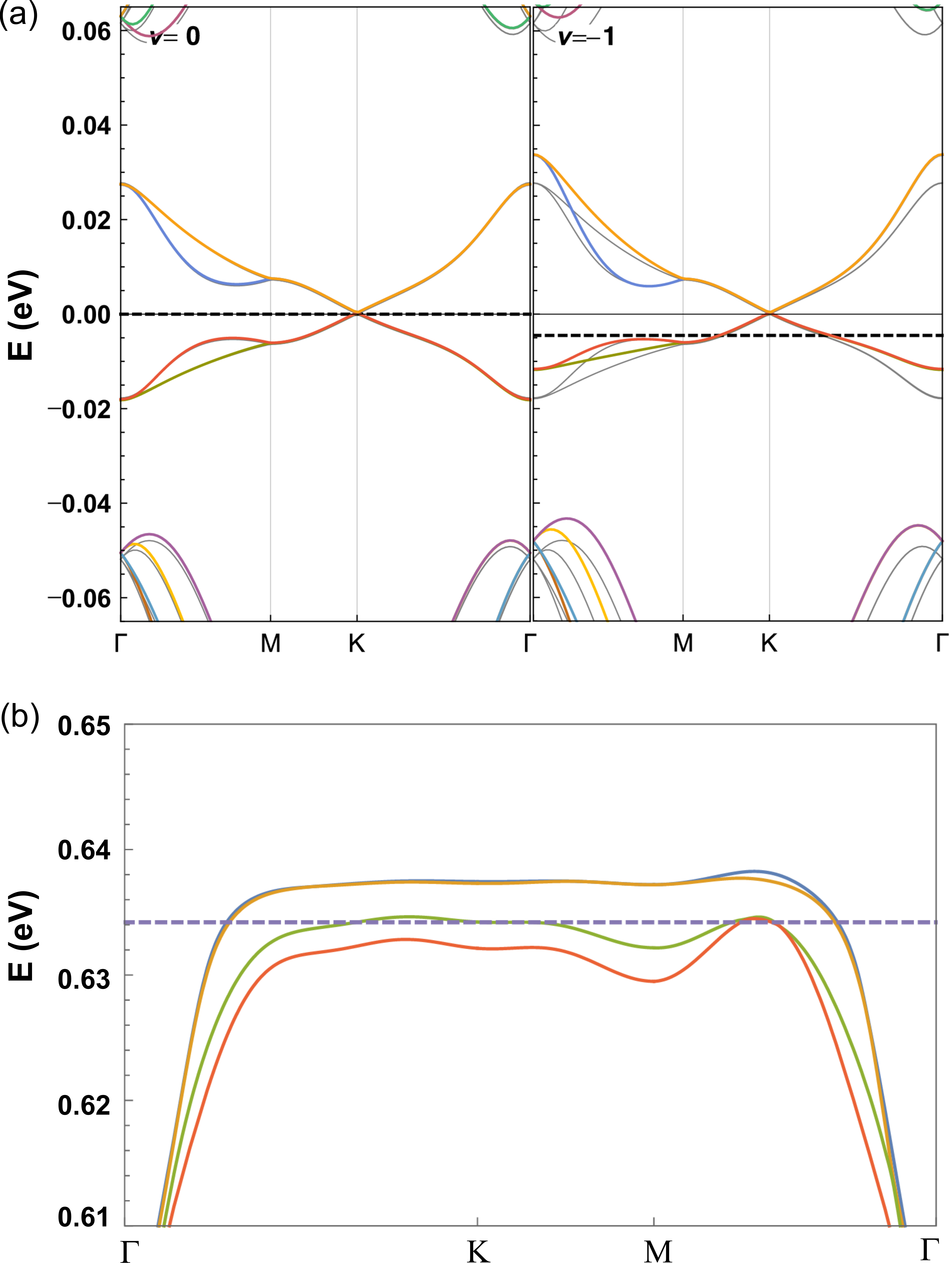}
 \caption{(a) The band structure of TBG with $\theta=1.08^\circ$ by taking long-range Hartree corrections into account at electron filling number $\nu=0$ (left side) and $\nu=-1$ (right side). The gray lines are the band structure without Hartree corrections. The dashed line is the Fermi level at each filling. Adapted with permission from \cite{rademaker2019charge}. Copyright (2019) by the American Physical Society. (b) The first valence (upper) and conduction (lower) flat bands obtained by including the Hartree-Fock interaction at filling number $\nu=0$ for TBG with $\theta=1.16^\circ$. The dashed line is the Fermi level. Adapted with permission from \cite{gonzalez2020time}. Copyright (2020) by the American Physical Society.}
 \label{figure:HF}
\end{figure}

\subsubsection{TB with electronic interactions}

The localization of electrons in flat bands near the Fermi energy results in strong electronic interactions that cannot be ignored in graphene based moiré materials. For the long range electron-electron interactions, the simplest model is the Hartree approximation, which accounts for a mean field direct interaction between an electron and the surrounding charge density. In TBG, this interaction has been found to be strongest near the magic angle and can be incorporated into the single particle TB model as~\cite{rademaker2019charge,goodwin2020hartree,fischer2022unconventional,cheung2022atomistic}
\begin{equation}
H = H_0 + H_H,
\label{eq:H_0H_H}
\end{equation}
where
\begin{equation}
H_H = \sum_i \delta n(\mathbf{r}_i) \phi_i
\end{equation}
is a self consistent Hartree potential. The electron interaction is replaced by a site-dependent electric potential $\phi_i$, which is determined self consistently through the equation
\begin{equation}
\phi_i = \sum_j V(\mathbf{r}_i - \mathbf{r}_j) \langle \delta n(\mathbf{r}_j) \rangle,
\label{eq:Hartree_energy}
\end{equation}
where $\delta n(\mathbf{r}) \equiv n(\mathbf{r}) - \bar{n}$ is the deviation of the electron density $n(\mathbf{r})$ from the average density $\bar{n}$, and $V(\mathbf{r}_i - \mathbf{r}_j)$ is the screened Coulomb interaction. The simplest form of this interaction can be written as~\cite{rademaker2019charge}
\begin{equation}
V(\mathbf{r}_i - \mathbf{r}_j) = \frac{1.438}{0.116 + \lvert \mathbf{r}_i - \mathbf{r}j \rvert} \, \text{eV}, \label{eq:Vij}
\end{equation}
but this potential can take different forms depending on the surrounding environment~\cite{goodwin2020hartree}, which has important effects when calculating the electronic interactions. Equations~\eqref{eq:H_0H_H} to~\eqref{eq:Vij} define a self consistent iterative scheme to obtain the band structure and eigenstates of the system. From these equations we can deduce the electronic density and then compute the electric potentials $\phi_i$. The electronic density can be expressed in terms of the Bloch eigenstates $\psi_{nk}(\mathbf{r})$ (with $n$ the band index and $\mathbf{k}$ the crystal momentum) of the Hamiltonian in Eq.\eqref{eq:H_0H_H} as
\begin{equation}
n(\mathbf{r}) = \sum_{nk} f_{nk} |\psi_{nk}(\mathbf{r})|^2,
\end{equation}
where $f_{nk} = 2\Theta(\varepsilon_F - \varepsilon_{nk})$ is the occupancy at zero temperature of the state $\psi_{nk}$ with eigenvalue $\varepsilon_{nk}$, $\varepsilon_F$ is the Fermi energy, and $\Theta(\varepsilon)$ is the Heaviside step function. In Fig.~\ref{figure:HF} we show the results of Ref.~\cite{gonzalez2020time}, where a TB model with a Hartree potential gives filling dependent renormalized flat bands near the Fermi energy. Similar results are obtained in Ref.~\cite{rademaker2019charge}. The TB results are also consistent with those from continuum models including the Hartree potential~\cite{guinea2018electrostatic,cea2019electronic,Cea2020BandCoulomb}.

To go beyond the Hartree approximation one can consider the Fock contribution, which accounts for the non local electronic interaction. The Fock approximation can be seen as the simplest effective description of the exchange interaction of electrons. Together with the Hartree interaction, this gives the mean field Hartree-Fock approximation. An example of a Hamiltonian with electron-electron interactions in twisted bilayer and trilayer graphene is a mean field Hartree-Fock Hamiltonian of the form~\cite{sanchez2024nematic,sanchez2024correlated,gonzalez2020time,gonzalez2021magnetic,gonzalez2023ising,sanchez2025nonflat}
\begin{eqnarray}
H_{MF} &=& H_0 + H_{HF} \\ \nonumber
&=& H_0+\sum_{\substack{i \neq j, s, s'}} V(\mathbf{r}_i - \mathbf{r}_j)  \langle c_{is}^{\dagger} c_{is} \rangle_0 c_{is'}^{\dagger} c_{is'} \\ 
&-&\sum_{\substack{i \neq j,s}} V(\mathbf{r}_i - \mathbf{r}_j') \langle c_{js}^{\dagger} c_{is} \rangle_0 c_{is}^{\dagger} c_{js}, \nonumber
\label{eq:hf}
\end{eqnarray}
where $H_0$ is the spin independent non interacting Hamiltonian of Eq.~\eqref{hal0}, $s(s')$ is the spin quantum number, which can be ignored when considering spin symmetric solutions~\cite{sanchez2024correlated}, and $\langle \cdots \rangle_0$ denotes the expectation value in a reference state. This HF equation $H_{\text{MF}}$ can be solved self consistently~\cite{sanchez2024nematic,sanchez2024correlated,gonzalez2020time,gonzalez2021magnetic,gonzalez2023ising,sanchez2025nonflat}. We note that the Hartree-Fock solution predicts a gap opening at the Dirac points, as shown in Fig.~\ref{figure:HF}(b), a result that is also captured by low energy continuum models\cite{Cea2020BandCoulomb}.

\begin{figure}
     \centering
     \includegraphics[width=0.5\textwidth]{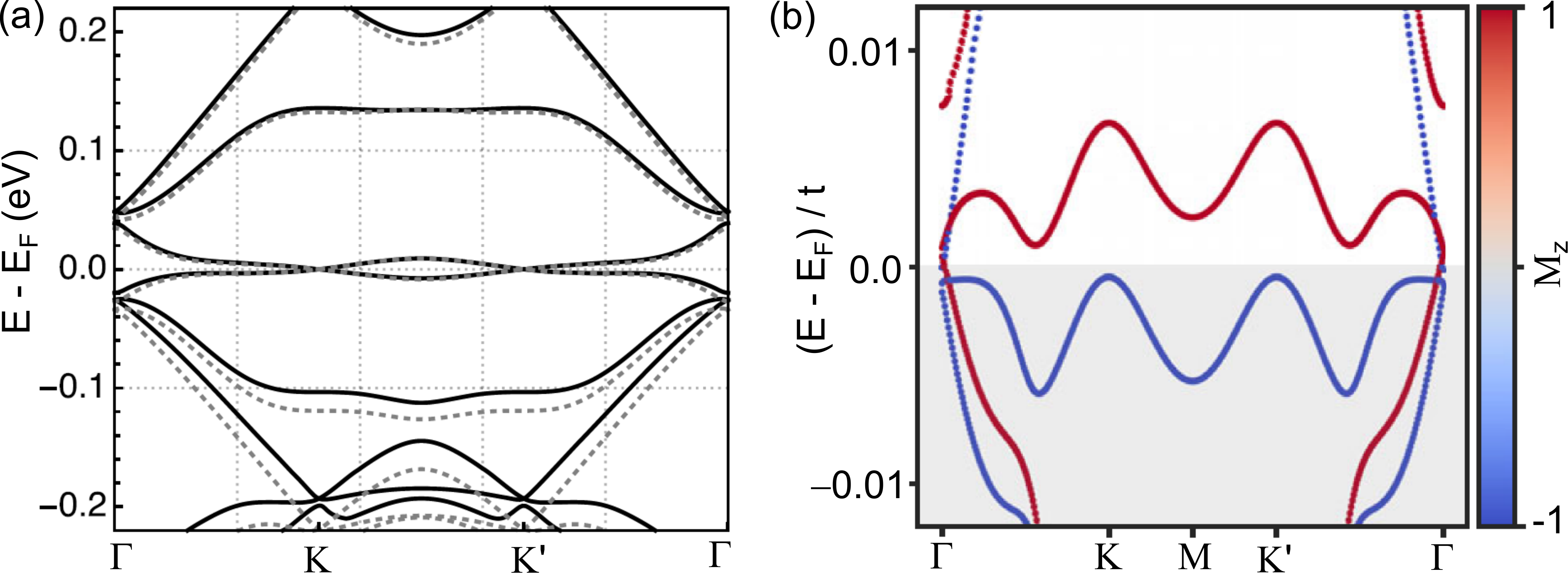}
     \caption{(a) Band structure of TBG with $\theta=1.5^\circ$ obtained from a scaled (solid lines) and unscaled (dashed lines) TB models. Adapted with permission from \cite{gonzalez2017electrically}. Copyright (2017) by the American Physical Society. (b) The calculated flat bands and spin $z$ magnetization of TBG with angle $\theta = 0.8^\circ$ by considering the effect of local mean-field interactions. At angle $\theta = 0.8^\circ$, the second bands from both conduction and valence bands became flat. The interaction strength is $U=2t$, where $t$ the nearest-neighbor hopping within one layer, and the electron filling number is $\nu = -6$, corresponding to half-filling of the second band. The red (blue) color indicates a positive (negative) expectation value $\langle S_z \rangle=M_z$ of the spin operator. The calculation was performed by using a rescaling method. Adapted with permission from \cite{wolf2019electrically}. Copyright (2019) by the American Physical Society.}
     \label{figure:hubbard}
\end{figure}

\subsubsection{TB model with Hubbard-U interaction} 
In moir\'e superlattices, the localized states of flat bands imply strong local electron-electron interaction that could lead to Mott insulating states, ferromagnetism \cite{cao2018correlated,tang2020simulation,chen2020tunable} and other correlated phases~\cite{JimenoPozo2023Short}. This short range interaction can be described in a minimal way using a local Hubbard term
\begin{equation}
    H_U = U\sum_i n_{i\uparrow}n_{i\downarrow},
\end{equation}
where $n_{i\uparrow} $($n_{i\downarrow}$) is the electron density operator $c_{i\uparrow}^{\dagger} c_{i\downarrow}$ ($c_{i\downarrow}^{\dagger} c_{i\uparrow}$) at each site for $p_z$ orbital and $U$ is the interaction strength. In a mean-field approximation, the TB Hamiltonian with the Hubbard-U term can be expressed as \cite{gonzalez2017electrically,wolf2019electrically} 
\begin{equation}
H_{MF} = H_0 + H_U \approx H_0 + U\sum_i \langle n_{i \uparrow} \rangle n_{i \downarrow} + n_{i \uparrow} \langle n_{i \downarrow} \rangle - \langle n_{i \uparrow} \rangle \langle n_{i \downarrow} \rangle,
\end{equation}
where $H_0$ is a single-particle TB Hamiltonian. The mean-field values $\langle n_{i \uparrow,\downarrow} \rangle$ are obtained by iteration until convergence. For TBG around the magic angle, the self-consistent process is typically time-consuming due to the large number of atoms in each moir\'e unit cell. Therefore, a rescaled non-interacting TB Hamiltonian is proposed to reach an affordable numerical self-consistent calculation. Specially, the low-energy electronic structure of TBG with a small angle $\theta$ can be reproduced at a larger angle $\theta'$ that contains a smaller number of atoms \cite{gonzalez2017electrically,wolf2019electrically}. The rescaled Hamiltonian can be obtained by tuning the parameters in Eqs. (\ref{eq:vpppi}) and (\ref{eq:vppsigma}) by the following scaling transformations~\cite{gonzalez2017electrically}
\begin{equation}
t_0' \rightarrow \frac{1}{\lambda} t_0, \;  
d' \rightarrow \lambda d,\; h' \rightarrow \lambda h, 
\end{equation}
where the dimensionless re-scaling parameter $\lambda$ is given by
\begin{equation}
\lambda = \frac{\sin \frac{\theta'}{2}}{\sin \frac{\theta}{2}}. 
\end{equation}
Figure~\ref{figure:hubbard}(a) shows the band structure of TBG with $\theta=1.5^{\circ}$ obtained from a scaled (solid lines) and an unscaled (dashed lines) TB Hamiltonian. The two methods give bands that agree well in the low energy region. The rescaled TB model with a mean field Hubbard $U$ Hamiltonian at the atomic level provides insight into ferromagnetism and Mott insulating states in TBG and other moir{\'e} superlattices~\cite{gonzalez2017electrically,wolf2019electrically,vahedi2021magnetism}. As shown in Fig.~\ref{figure:hubbard}(b), at half filling of the second band, interactions induce a Stoner instability that splits the flat bands\cite{wolf2019electrically}.

\subsection{TMDs-based moir\'e materials}

Another family of materials for moir\'e physics is TMDs, which have attracted growing interest in condensed matter physics. Recently, exciting experimental phenomena, like moiré flat bands, correlated insulating states, interfacial ferroelectricity, Wigner crystals, superconductivity, have been observed in twisted TMDs \cite{zhang2020flat,xu2020correlated,weston2022interfacial,regan2020mott,guo2025superconductivity,xia2025superconductivity}. TMDs have a triangular geometry that can host both hexagonal (2H) and tetragonal (1T) stackings. In monolayer TMDs, each cell contains one metal and two chalcogenide elements with chemical formula MX$_2$. The geometry and electronic properties vary with different elements. Interestingly, the bilayer moir\'e pattern can be generated by both identical monolayers (homobilayer) and different monolayers (heterobilayer). In the following, we will describe the TB model for the TMDs homobilayer and heterobilayer.

\subsubsection{TB for twisted homobilayer TMDs}

In general there are three TB models for the homobilayer TMDs moir\'e systems. These three TB models propose very different parameters (onsite energies and SK parameters), but provide electronic structures that are highly consistent. All models adapt an 11-orbital in the monolayer, but consider different interlayer interactions.  One of the TB models, discussed by Zhan and coworkers \cite{zhan2020tunability,zhang2020tuning}, considers only the interlayer interactions between the $p$ orbitals of the $X$ atoms at the interface between the two layers. The corresponding TB parameters were developed by Fang and coworkers \cite{fang2015ab}. In the following we describe the theory of this TB model.

The geometry of the TMDs moir\'e patterns can be defined in the same manner of the graphene moir\'e systems. The bilayer TB Hamiltonian can be derived by adding an interlayer hopping term to two monolayer Hamiltonians as 
\begin{equation}
	\hat H =\hat H_1^{(1L)}+\hat H_2^{(1L)}+ \hat H_{int}^{(2L)},
	\label{eq:hal_twtmd}
\end{equation}
where the first two terms are the monolayer Hamiltonians and the third term is the interlayer hopping term. The monolayer TB model is constructed from an 11 basis set (five $d$ orbitals from $M$ and three $p$ orbitals from $X$) as $\hat{\psi}_{pd}^{\dagger} =
[\hat{d}_z^{\dagger}, \hat{d}_{xy}^{\dagger}, \hat{d}_{x^2 - y^2}^{\dagger}, \hat{d}_{xz}^{\dagger}, \hat{d}_{yz}^{\dagger},
\hat{p}_x^{A \dagger}, \hat{p}_y^{A \dagger}, \hat{p}_z^{A \dagger}, \hat{p}_x^{B \dagger}, \hat{p}_y^{B \dagger}, \hat{p}_z^{B \dagger}]$, which contains the on-site energy, the hopping terms between orbitals of the same type at first-neighbor positions, and the hopping terms between orbitals of different type at first- and second-neighbor positions \cite{fang2015ab}. TB parameters in the single-layer Hamiltonian for MoS$_2$, MoSe$_2$, WS$_2$ and WSe$_2$ can be obtained from Table VII of Ref. \cite{fang2015ab}. The term $\hat H_{int}^{(2L)}$ is the interlayer interaction expressed as 
\begin{eqnarray}
H_{int}^{2L} = \displaystyle\sum_{p_i',\mathbf r_2,p_j,\mathbf r_1}\hat \phi_{2,p_i'}^\dagger(\mathbf r_2)t_{p_i',p_j}^{(LL)}(\mathbf r_2-\mathbf r_1)\hat \phi_{1,p_j}(\mathbf r_1) + \; \mathrm{H. c.}, 
\end{eqnarray}
where $\hat \phi_{i,p_j}$ is the $p_j$ orbital basis of $i$-th monolayer. Within the SK parametrization, the interlayer hoppings are expressed as\cite{SK1954simplified}
\begin{equation}
	t_{p_i',p_j}^{(LL)}(\mathbf r) = (V_{pp,\sigma}(r)-V_{pp,\pi}(r))\frac{r_ir_j}{r^2}+V_{pp,\pi}(r)\delta_{i,j},
\end{equation}
where $r=|\mathbf r|$ and the distance-dependent SK parameter is
\begin{equation}
	V_{pp,b}=\nu_be^{[-(r/R_b)^{\eta_b}]},
	\label{inter}
\end{equation}
where $b=\sigma,\pi$, $\nu_b$, $R_b$ and $\eta_b$ are constant values that can be obtained from the Ref. \cite{fang2015ab}. The interlayer interactions in twisted homobilayer TMDs are included in the TB Hamiltonian by adding hoppings between p orbitals of chalcogen atoms in top and bottom layers. The cuttoff distance of interlayer hopping can be taken as 5 \AA~\cite{zhang2020tuning,zhan2020tunability,kuang2022flat}. The TMDs have two set of bond length values (theoretical and experimental bulk values). For MoS$_2$, the bond lengths are the in-plane lattice constant $a=3.18\, [3.16]\,\text{\AA}$, unit cell size along the z direction $h=[12.29]\,\text{\AA}$, distance alone z direction between chalcogen layers $d_{X-X}=3.13\,[3.17]\,\text{\AA}$, nearest-neighbor bond betwwen metal and chalgogen atoms $d_{X-M}=2.41\,[2.42]\,\text{\AA}$\cite{fang2015ab}. Values in brackets are experimental bulk values. More details on the bond length of other TMDs refer to Refs. \cite{fang2015ab,cappelluti2013mos2}.    

Strong spin-orbital coupling (SOC) is a main characteristic in TMDs. By expanding the 11 orbitals to 22, SOC can be incorporated into the TB model. The intralayer Hamiltonian of Eq. (\ref{eq:hal_twtmd}) with SOC is given by \cite{fang2015ab}
\begin{equation}
\begin{split}
\hat{H}_{\text{SO}}^{(1L)} &= \sum_{\mathbf{k}} \big[ 
\hat{\boldsymbol{\phi}}_{\uparrow}^{\dagger}(\mathbf{k}) H_{\uparrow\uparrow}^{(1L)}(\mathbf{k}) \hat{\boldsymbol{\phi}}_{\uparrow}(\mathbf{k})\\
&+ \hat{\boldsymbol{\phi}}_{\downarrow}^{\dagger}(\mathbf{k}) H_{\downarrow\downarrow}^{(1L)}(\mathbf{k}) \hat{\boldsymbol{\phi}}_{\downarrow}(\mathbf{k})
+ \hat{\boldsymbol{\phi}}^{\dagger}(\mathbf{k}) H_{\text{LS}} \hat{\boldsymbol{\phi}}(\mathbf{k})
\big].\\
\end{split}
\end{equation}
The diagonal blocks in the first term \( H_{\uparrow\uparrow}^{(1L)} = H_{\downarrow\downarrow}^{(1L)} = H^{(1L)} \) 
are the intralayer Hamiltonian. These are the spin-independent hopping processes. The effect of spin-orbit coupling, \( H_{\text{LS}} \), 
is incorporated by the on-site \( \lambda_{\text{SO}} \mathbf{L} \cdot \mathbf{S} \) term for each atom. Because it is an on-site term, it does not 
carry momentum dependence and is a constant matrix with elements
\begin{equation}
\langle \phi_{i,\sigma} | H_{\text{LS}} | \phi_{j,\sigma'} \rangle
= \langle \phi_{i,\sigma} | \big( \lambda_{\text{SO}}^M \mathbf{L}_M + \lambda_{\text{SO}}^X \mathbf{L}_X^A + \lambda_{\text{SO}}^X \mathbf{L}_X^B \big) \cdot \mathbf{S} | \phi_{j,\sigma'} \rangle,
\end{equation}
where $\lambda_{\text{SO}}^M$ and $\lambda_{\text{SO}}^X$ are the SOC strength of the $M$ and $X$ atoms, respectively\cite{fang2015ab}. Within SOC, the interlayer Hamiltonian will only consider the interaction of electrons with the same spin direction. In this way, the tunable SOC in twisted homobilayer and homotrilayer TMDs were carefully studied \cite{zhan2020tunability,li2024tuning}.

Lattice relaxation is also an important effect in TMDs moir\'e systems that needs to be taken into account in the TB model. When relaxing the system, atoms moves away from its equilibrium position, both in-plane and out-of-plane. Upon relaxation, the intralayer hoppings can be modified through the form\cite{rostami2015theory}
\begin{equation}
	t_{ij,\mu\nu}^{intra}(\mathbf r_{ij})=t_{ij,\mu\nu}^{intra}(\mathbf r_{ij}^0)\bigg(1-\Lambda_{ij,\mu\nu}\frac{|\mathbf r_{ij}-\mathbf r_{ij}^0|}{|\mathbf r_{ij}^0|}\bigg),
\end{equation}
where $t_{ij,\mu\nu}^{intra}$ is the intralayer hopping between the $\mu$ orbital of the $i$ atom and $\nu$ orbital of the $j$ atom, $\mathbf r_{ij}^0$ and $\mathbf r_{ij}$ are the distance between the $i$ and $j$ atoms in the equilibrium and relaxed cases, and $\Lambda_{ij,\mu\nu}$ is the dimensionless bond-resolved local electron-phonon coupling. It is assumed that $\Lambda_{ij,\mu\nu}=3,4,5$ for the chalcogen-chalcogen $pp$, chalcogen-metal $pd$ and metal-metal $dd$ hybridizations, respectively\cite{rostami2015theory}. By using the TB model, ultraflat bands were found to exist in TMDs for almost any small twist angles \cite{zhan2020tunability}.

The second TB model was presented by Venkateswarlu and coworkers \cite{venkateswarlu2020electronic}. In this TB model, the interlayer interaction included $p$\;S--$p$\;S, $d$\;Mo--$p$\;S and $d$\;Mo--$d$\;Mo  terms. The TB parameters were set up to correctly match the DFT band structures. 

\begin{figure}
     \centering
     \includegraphics[width=0.45\textwidth]{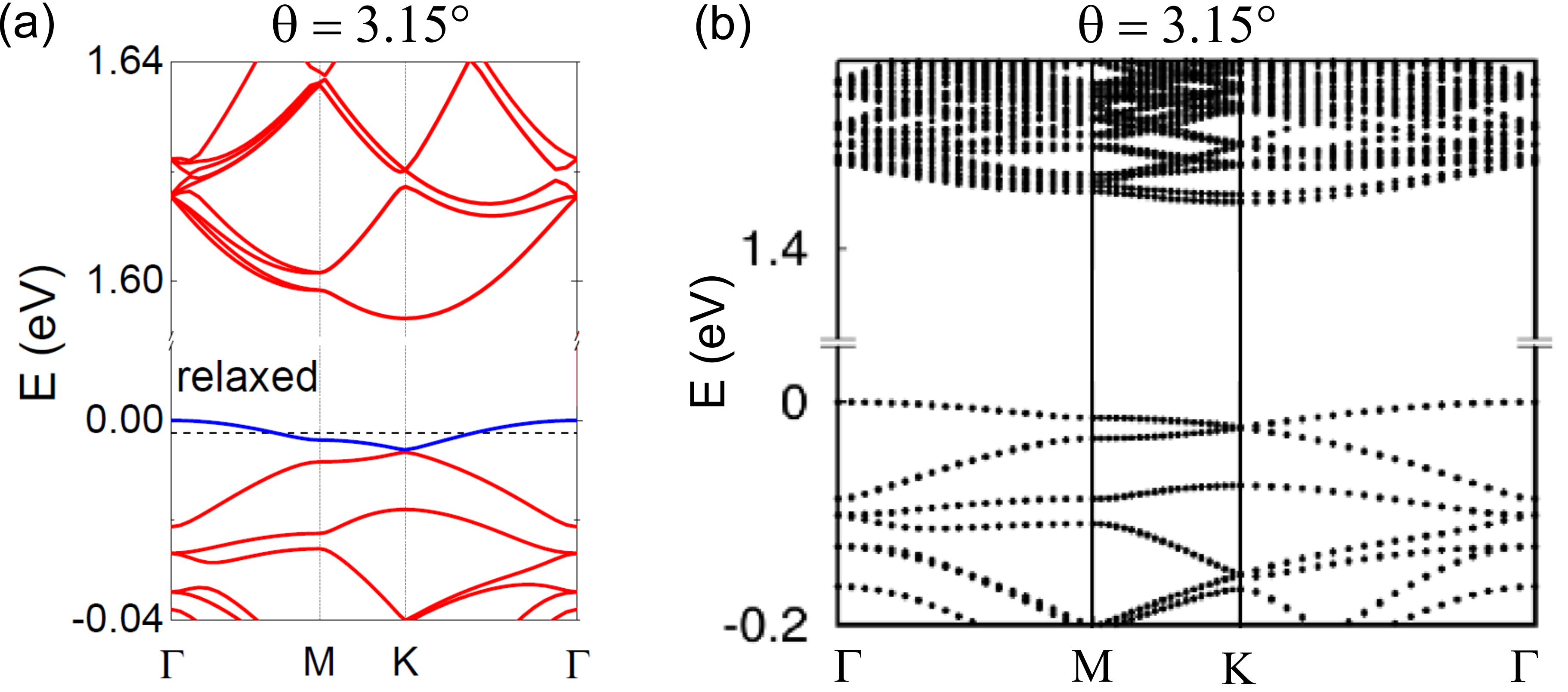}
     \caption{Tight-binding band structure of twisted homobilayer MoS$_2$ at $\theta = 3.15^\circ$. (a) Bands obtained from a TB model from Ref. \cite{zhan2020tunability,kuang2022flat}. Adapted with permission from \cite{kuang2022flat}. Copyright (2022) by the American Physical Society. (b) Bands calculated from the TB model from Ref. \cite{vitale2021flat}. Adapted under the terms of the CC BY license from Ref. \cite{vitale2021flat}. Copyright (2021) IOP Publishing.}
     \label{figure:twhomotmd}
\end{figure}

In the third TB model, formulated by Vitale and coworkers~\cite{vitale2021flat}, the interlayer interactions $p$;S--$p$;S and $p_z$;S--$d_{z^2}$;Mo were included. Moreover, they described the interlayer hoppings ($p$--$p$ and $p_z$--$d_{z^2}$) using different sets of SK parameters for varying interlayer separations. The TB parameters were obtained from a Wannier transformation of the DFT Hamiltonian. Figure~\ref{figure:twhomotmd} shows the band structures of twisted MoS$_2$ with the same twist angle but derived from different TB models. The results are highly consistent with one another.

\subsubsection{TB for twisted heterobilayer TMDs} 

\begin{figure}
     \centering
     \includegraphics[width=0.5\textwidth]{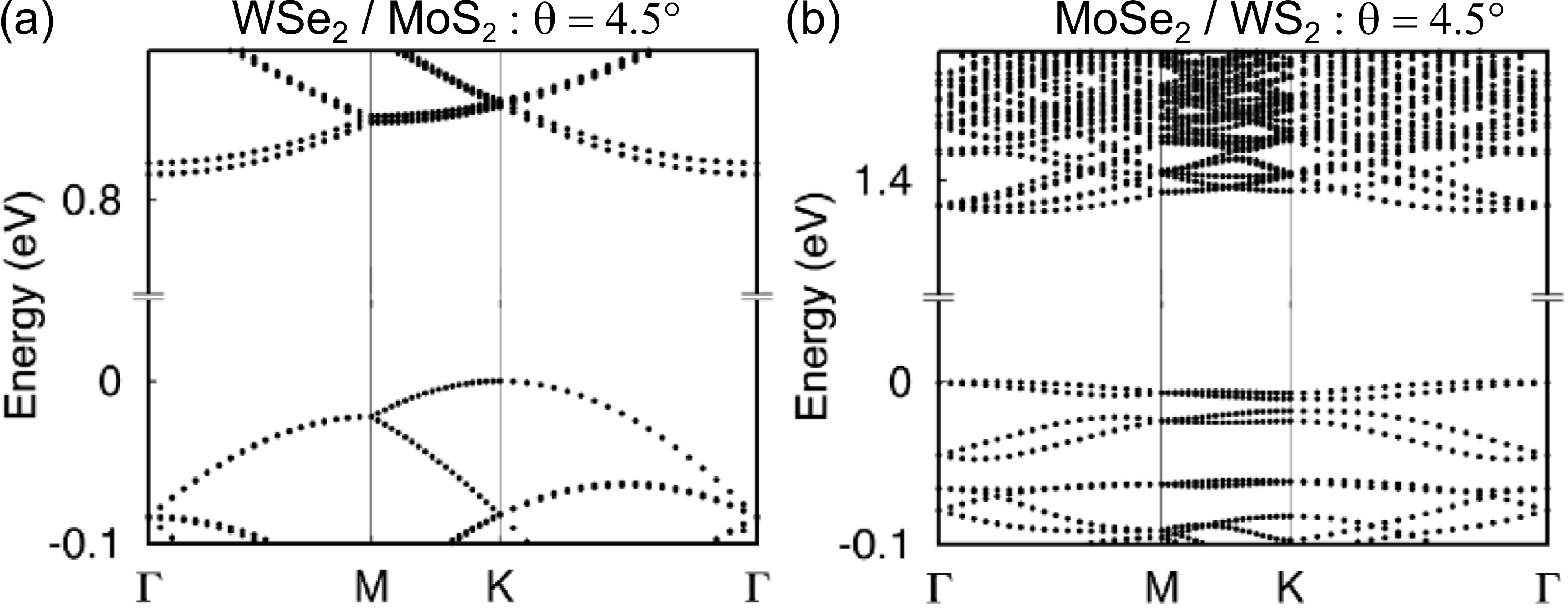}
     \caption{Tight-binding band structure of twisted heterobilayer TMDs for (a) twisted bilayer WSe$_2$/MoS$_2$ and (b) MoSe$_2$/WS$_2$ heterostructure at twist angle $\theta = 4.5^\circ$\cite{vitale2021flat}. Adapted under the terms of the CC BY license from Ref. \cite{vitale2021flat}. Copyright (2021) IOP Publishing.}
     \label{figure:heter_TMDC}
\end{figure}

In 2021, by fitting DFT band strutures, Vitale and coworkers extended the work of Fang and \textit{et al.}, to construct the TB Hamiltonian for both twisted heterobilayer and homobilayer TMDs. In this TB model, they also consider the interlayer hoppings between chalcogen $p$ and metal $d_{z^2}$ orbitals with a SK expression\cite{vitale2021flat}
\begin{equation}
t_{p_z, d_{z^2}} (\mathbf{r}) = n \left[ n^2 - \frac{1}{2} (l^2 + m^2) \right] V_{pd\sigma} (\mathbf{r}) + \sqrt{3} n (l^2 + m^2) V_{pd\pi} (\mathbf{r}),
\end{equation}
where the directional cosines are defined as \( l = r_x / r \), 
\( m = r_y / r \) and \( n = r_z / r \). To determine the functions \( V_{pd\sigma} (\mathbf{r}) \) and \( V_{pd\pi} (\mathbf{r}) \), 
Vitale and coworkers calculated \( t_{p_z, d_{z^2}} \), \( t_{p_z, d_{xz}} \) and \( t_{p_z, d_{yz}} \) 
for a set of untwisted bilayers with different stacking configurations 
and different interlayer separations, using a Wannier transformation of the DFT Hamiltonian. Then, a least square fitting process was used to extract \( V_{pd\sigma} \) 
and \( V_{pd\pi} \) at different interatomic distances. The results 
were fitted to functions of the type
\begin{equation}
V_{pd,b}(\mathbf{r}) = V_b \left( \frac{r}{h} \right)^{\alpha_b} 
\cos \left( \beta_b \frac{r}{h} + \gamma_b \right),
\end{equation}
where $b=\sigma, \pi$ \( V_b, \alpha_b, \beta_b \) and \( \gamma_b \) denote interlayer hopping parameters fitted from DFT calculations, which are dependent on the types of heterostructures of bilayer TMDs\cite{vitale2021flat}. \( h = 3.5 \) \AA\ is an average interlayer distance. All the TB parameters are in Ref. \cite{vitale2021flat}. Figure \ref{figure:heter_TMDC} shows the band structures of the TMDs heterostructures containing different species of chalcogens. Similar to the homobilayer case, the highest valence bands are derived from monolayer $K/K'$ states (Fig. \ref{figure:heter_TMDC}(a)) or $\Gamma$ states (Fig. \ref{figure:heter_TMDC}(b)).   

\begin{table*}[t]
\label{table:TB}
\small
\setlength{\tabcolsep}{3pt}\renewcommand{\arraystretch}{1.12}

\begin{tabular*}{\textwidth}{@{\extracolsep{\fill}}%
  >{\raggedright\arraybackslash}p{0.11\textwidth}%
  >{\raggedright\arraybackslash}p{0.14\textwidth}%
  >{\raggedright\arraybackslash}p{0.28\textwidth}%
  >{\raggedright\arraybackslash}p{0.28\textwidth}%
  >{\raggedright\arraybackslash}p{0.12\textwidth}@{}}
\hline
\textbf{Material family} & \textbf{Orbitals (basis)} &
\textbf{Hopping used (intralayer / interlayer)} &
\textbf{Parameters (SK-related numerics)} &
\textbf{Relaxation (LAMMPS potentials)} \\
\hline
Moir\'e graphene (TBG \& stacks)
& $p_z$ per C
& \textit{intra/inter}\cite{trambly2010localization,trambly2012numerical}:
$t_{ij}=n^2 V_{pp\sigma}(r)+(1-n^2)V_{pp\pi}(r)$; $V_{pp\pi}(r)=-t_0 e^{q_\pi(1-r/d)} F_c(r)$; $V_{pp\sigma}(r)=t_1 e^{q_\sigma(1-r/h)} F_c(r)$; $F_c(r)=\big(1+e^{(r-r_c)/l_c}\big)^{-1}$
& $t_0=2.7\,\mathrm{eV}$; $t_1=0.48\,\mathrm{eV}$; $q_\sigma/h=q_\pi/d=2.218\,\mathrm{\AA^{-1}}$; 

$r_c=5.0\,\mathrm{\AA}$;

$l_c=0.265\,\mathrm{\AA}$
& \textit{intra}\cite{guinea2019continuum,kuang2021collective}: AIREBO\cite{stuart2000airebo}
LCBOP\cite{los2003lcbop};
\textit{inter}\cite{guinea2019continuum,leconte2022relaxation}:

Kolmogorov--Crespi(KC)\cite{kolmogorov2005registry}  \\
[2pt]
Moir\'e hBN (twisted bilayer)
& $p_z$ on B/N
& \textit{Intra}: (A)\cite{walet2021flat} and (B)\cite{sponza2024emergence} nearest neighbor (NN) hopping.
(c) 6 neighbor hoppings\cite{li2024moire}.
 \textit{Inter}: (A)\cite{walet2021flat} $t^{XY}_\perp(r)=t^{XY} e^{-\alpha(r-h)}$; (B)\cite{sponza2024emergence} $t^{XY}_\perp(r)=n^2 \gamma_{XY} F_c^{XY}(r)\, e^{Q_{XY}(h-r)}$; (C)\cite{li2024moire} full SK: $V_{pp\pi}(r)=-\gamma_0 e^{q_\pi(1-r/d_{BN})}$, $V_{pp\sigma}(r)=\gamma_1 e^{q_\sigma(1-r/h)}$
& (A)\cite{walet2021flat} $h=3.33\,\mathrm{\AA}$, $\alpha=4.4\,\mathrm{\AA^{-1}}$; $t_{NN}=0.15$, $t_{BB}=0.7$, $t_{NB}=0.3$ eV. 

(B)\cite{sponza2024emergence} $h=3.22\,\mathrm{\AA}$, $l_c=0.265\,\mathrm{\AA}$, $r_c^{XY}=h+\ln(10^3)/Q_{XY}$.

(C)\cite{li2024moire} $d_{BN}=1.43\,\mathrm{\AA}$,
$h=3.261\,\mathrm{\AA}$; $\gamma_0=2.7\,\mathrm{eV}$; $\gamma_1\in\{0.831,0.6602,0.3989\}\,\mathrm{eV}$
&  \textit{intra}\cite{li2024moire}: 
extended Tersoff\cite{los2017bntersoff};
\textit{inter}\cite{li2024moire}:
DRIP\cite{wen2018dihedral,jung2024commensuration} \\
[2pt]
Moir\'e TMDs (twisted homo/hetero)
& 11-orbital: $5d$ (M) + $p_{x,y,z}$ on two X; SOC on-site
& \textit{Intra}: Wannier 11-orbital TB\cite{fang2015ab}. \textit{Inter} (homobilayer $p$–$p$)\cite{fang2015ab}: $t^{(LL)}_{p'_i,p_j}(r)=(V_{pp,\sigma}-V_{pp,\pi})\frac{r_i r_j}{r^2}+V_{pp,\pi}\delta_{ij}$ with $V_{pp,b}(r)=\nu_b e^{-(r/R_b)^{\eta_b}}$. \textit{Inter} (heterobilayer $p_z$–$d_{z^2}$)\cite{vitale2021flat}: $t_{p_z,d_{z^2}}(r)=n\!\left[\frac{n^2-1}{2}(l^2+m^2)\right]\!V_{pd\sigma}(r)+\sqrt{3}\,n(l^2+m^2)V_{pd\pi}(r)$; $V_{pd,b}(r)=V_b(r/h)^{\alpha_b}\cos(\beta_b r/h+\gamma_b)$
& Homobilayer\cite{zhang2020flat}: 

interlayer cutoff $r_{\mathrm{cut}}\!\approx\!5\,\mathrm{\AA}$; $\nu_b$, $R_b$, $\eta_b$ from Table V in ~\cite{fang2015ab}.

Heterobilayer\cite{vitale2021flat}:

mean interlayer distance $h\!\approx\!3.5\,\mathrm{\AA}$; 

$(V_b,\alpha_b,\beta_b,\gamma_b)$ from \cite{vitale2021flat} 
& \textit{intra}\cite{vitale2021flat,kuang2022flat}: 
Stillinger--Weber (SW)\cite{jiang2017sw2d};
\textit{inter}\cite{kuang2022flat,vitale2021flat}:
Lennard-Jones (LJ) \cite{rappe1992uff}/KC
\cite{naik2019kolmogorov} \\
\hline
\end{tabular*}
\caption{Summary of tight-binding (TB) models that have been used for moiré superlattices of three representative material families: (i) twisted bilayer and multilayer graphene, (ii) twisted bilayer hexagonal boron nitride (hBN), and (iii) twisted homobilayer and heterobilayer transition–metal dichalcogenides (TMDs). 
The second column lists the orbital basis actually used in the TB Hamiltonians (from the simplest $p_z$ model for graphene to the 11-orbital Wannier model for TMDs). 
The third column specifies the intralayer and interlayer hopping functions, including SK parametrizations, range cutoffs, and angle dependences reported in the cited works. 
The fourth column collects SK-related numerical parameters (lattice constants, onsite energies, hopping amplitudes and decay lengths).
The last column summarizes how the atomic structures were relaxed via LAMMPS with corresponding intra- and inter-layer potetials before evaluating SK matrix elements }
\end{table*}

\subsection{hBN-based moir\'e materials}
\subsubsection{TB for twisted bilayer hBN}

Similar to TMDs, the bilayer hBN has two possible distinct stacking configurations, the parallel BN/BN and antiparallel alignment BN/NB. In the beginning, the twisted bilayer hBN was studied by DFT calculations, unveiling multi-flat bands at the edges of the bands at an angle $\theta=2.64^{\circ}$, and no constraint of magic angles that was similar to the TMDs case\cite{xian2019multiflat}. Therefore, twisted bilayer hBN could provide an ideal platform to study correlations effects. However, the DFT calculations could only tackle large angle systems. Thus, an atomic TB model was proposed by Walet and Guinea, which could further facilitate finer studies of electronic properties for small angle twisted bilayer hBN\cite{walet2021flat}. In this TB model, the twisted bilayer hBN Hamiltonian is composed of intralayer $H_{1(2)}$ and interlayer $H_{12}$ parts
\begin{equation}
    H = H_1 + H_2 + H_{12}.
\end{equation}
$H_{1(2)}$ is similar to the single-layer Hamiltonian of graphene and has the form: 
\begin{equation}
	\hat{H}_{1(2)} = \sum_i \epsilon_i c_i^{\dagger} c_i^{\phantom{\dagger}} - \sum_{<i,j>} t c_i^{\dagger} c_j^{\phantom{\dagger}},
    \label{eq:hal_hbn_intra}
\end{equation}
in which $i$ denotes the $p_z$ orbital site of $B$ or $N$ atom. $\epsilon_i$ is the onsite energy that has a difference $\Delta = \epsilon_B - \epsilon_N$ for B and N atoms\cite{ribeiro2011stability}.  $t$ is the intralayer nearest hopping between B and N. $\Delta$ and $t$ are set as 8 eV and 2.33 eV, respectively \cite{walet2021flat}. $H_{12}$ is the interlayer Hamiltonian with the form\cite{walet2021flat}
\begin{equation}
t_{\perp}^{XY}(r) = t_{XY} \exp\bigl(-\alpha\,(r - h)\bigr),
\label{eq:interhbnwalt}
\end{equation}
where $r$ is the distance between X and Y atoms (X(Y) is B or N) and the empirical parameters are set as \( h = 3.33\,\text{\AA} \), \(\alpha = 4.4\,\text{\AA}^{-1}\), $t_{NN}=0.15$ eV, $t_{BB} = 0.7$ eV and $t_{NB} = 0.3$ eV in Ref. \cite{walet2021flat}. In the above TB model, the hopping term in Eq. (\ref{eq:interhbnwalt}) does not distinguish the atomic species, and assumes one distance-dependent relation for all atoms. However, this model could capture the flat band features and give an explanation of charge polarization in twisted bilayer hBN\cite{walet2021flat,woods2021charge}.

\begin{figure}[t]
     \centering
     \includegraphics[width=0.5\textwidth]{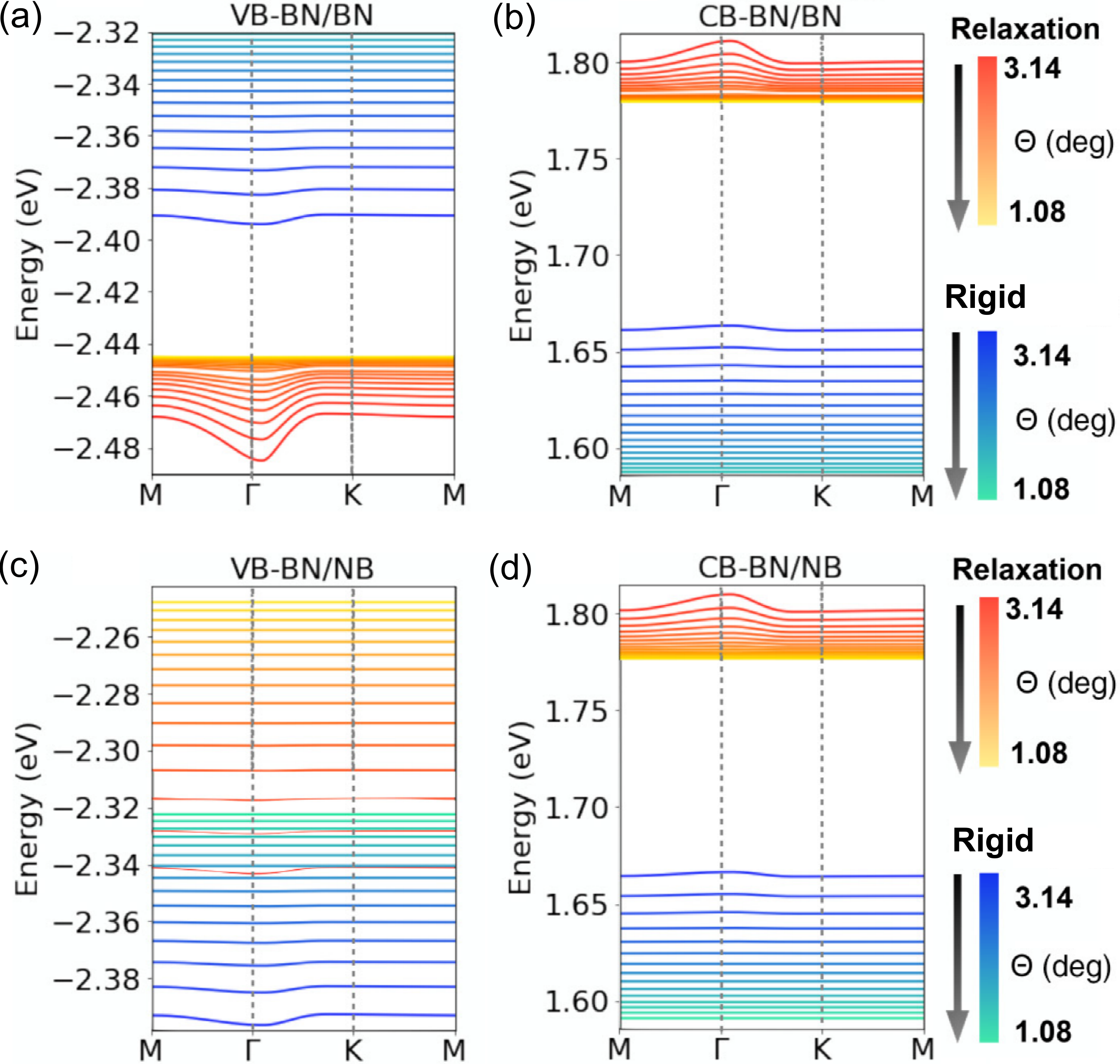}
     \caption{(a) Low-energy valence band and (b) conduction band for the BN/BN stacking with various twist angles, for rigid and relaxed configurations. (c) and (d) are the same plot but for BN/NB. Adapted with permission from \cite{li2024moire}. Copyright (2024) by the American Physical Society.}
     \label{figure:twist_hBN}
\end{figure}

Two additional TB models, fitted from DFT results for twisted bilayer hBN, have been proposed~\cite{li2024moire,sponza2024emergence}. One of them, developed by Sponza and coworkers, employs the first nearest-neighbor in-plane Hamiltonian [Eq.~(\ref{eq:hal_hbn_intra})] with $\epsilon_B = 4.90$ eV, $\epsilon_N = 0$ eV, and $t = 2.65$ eV, and uses a TBG-like relation that includes only $V_{pp\sigma}$ for the interlayer hopping~\cite{sponza2024emergence}.
\begin{equation}
t_\perp^{XY}(r) \;=\; n^2\,\gamma^{XY}\,F_c^{XY}(r)\,
\exp\Bigl[\,Q_{XY}\,\bigl(h - r\bigr)\Bigr],
\label{eq:t_sigma}
\end{equation}
where $h=3.22$ \AA\; is the interlayer distance, $XY$ labels the pairings BN, BB, or NN, and 
\begin{equation}
F_{c}^{XY}\left(r\right)=\frac{1}{1+\exp\left[\left(r-r_{c}^{XY}\right)/l_{c}\right]}
\end{equation}
is a smooth function with $l_c = 0.265\,\text{\AA}$ and cutoff distance $r_c^{XY}$. The values of $\gamma^{XY}$ and $Q_{XY}$ in Eq. (\ref{eq:t_sigma}) can be found in Ref. \cite{sponza2024emergence}. The cutoff distance $r_c^{XY}$ depends on the value of $Q_{XY}$ according to the relation $r_c^{XY} \;=\; h \;+\;\frac{\ln\bigl(10^3\bigr)}{Q_{XY}}$.

Another TB model, developed by Li and coworkers, considered intralayer hoppings up to six neighbors and used onsite energies of $\epsilon_B = 1.7666$ eV and $\epsilon_N = -2.1843$ for the first term in Eq.~(\ref{eq:hal_hbn_intra})~\cite{li2024moire}. In addition, the lattice relaxation effect could be incorporated into the intralayer interaction as
\begin{equation}
t_{\alpha\beta}(r_{ij})
= t_{\alpha\beta}(r_{0,ij}) 
\exp\!\Biggl[
  -2.45\,\biggl(\frac{r_{ij} - r_{0,ij}}{r_{0,ij}}\biggr)
\Biggr],
\end{equation}
where $t_{\alpha\beta}(r_{0,ij})$ is the intralayer hopping terms of the rigid lattice with distance $r_{0,ij}$ between atoms~$i$ and~$j$, and $r_{ij}$ is the relaxed distance. $r_{0,BB}$, $r_{0,BN}$ and $r_{0,NN}$ can be obtained by using the lattice constant $a = 2.4795$\,\AA\; of the rigid case. The interlayer hopping terms
are determined by the SK relation in Eq. (\ref{eq:tij}) with
\begin{align}
V_{pp\pi}(r_{ij}) &= -\gamma_{0}\,\exp\!\Bigl[
  q_{\pi}\,\bigl(1 - \tfrac{r_{ij}}{d_{\mathrm{BN}}}\bigr)
\Bigr], \nonumber\\
V_{pp\sigma}(r_{ij}) &= \gamma_{1}\,\exp\!\Bigl[
  q_{\sigma}\,\bigl(1 - \tfrac{r_{ij}}{h}\bigr)
\Bigr],
\label{eq:hbn_vppsigma}
\end{align}
where the intralayer distance is $d_{\mathrm{BN}}=a/\sqrt{3}=1.43$\,\AA,
the vertical interlayer distance is $h=3.261$\,\AA\,
, $\gamma_{0}=2.7$ eV, while $\gamma_1$ has $\gamma_{1} = t_{BB'} = 0.831$ eV, $\gamma_{1} = t_{NN'} = 0.6602$ eV, or $\gamma_{1} = t_{BN'} = t_{NB'} = 0.3989$ eV. The parameters $q_{\pi}$ and $q_{\sigma}$ have the relation
\begin{equation}
\frac{q_{\sigma}}{h}=\frac{q_{\pi}}{d_{\mathrm{BN}}}=\frac{\ln\left(0.1\gamma_{0}/\gamma_{0}\right)}{d_{\mathrm{BN}}-a}.
\end{equation} 
Figure \ref{figure:twist_hBN} shows the band structure of twisted bilayer hBN obtained from Li's TB Hamiltonian\cite{li2024moire}. The band gap increased significantly after lattice relaxation. The bands from the edges became extremely flat in the small-angle region.

\subsubsection{TB model for graphene/hBN moir\'e superlattice }

In experiments, hBN is widely used as a substrate to support or encapsulate graphene and twisted graphene layers. Its atomically flat surface and lack of dangling bonds improve the device quality by reducing disorder and enhancing carrier mobility. Because of the lattice mismatch between graphene and hBN, a graphene/hBN superlattice forms even when the lattices are crystallographically aligned. The presence of hBN modifies the electronic properties of graphene, multilayer graphene, and twisted graphene through interlayer interactions between carbon and B or N atoms. The total TB Hamiltonian can be written as
\begin{equation}
H = H_{g} + H_{hBN} + H_{\perp},
\end{equation}
where $H_{g}$ and $H_{hBN}$ denote the TB Hamiltonians of graphene and monolayer hBN, respectively. The single layer Hamiltonians $H_{g}$ and $H_{hBN}$ are as introduced in the previous sections. The key ingredient is the interlayer interaction $H_{\perp}$, which can be expressed using the Slater–Koster relation in Eq. (\ref{eq:tij}), with the same $V_{pp\pi}$ and $V_{pp\sigma}$ as in Eqs. (\ref{eq:vpppi}) and~(\ref{eq:vppsigma}). In most calculations, the hopping parameters $t_0$ and $t_1$ between a carbon atom and a B or N atom are set to $t_0 = 2.7$ eV and $t_1 = 0.48$ eV. A complementary route is to construct effective hBN potentials within TB models~\cite{moon2014electronic}. When lattice relaxation is important, combining atomistic TB with classical molecular dynamics provides a practical way to include structural relaxation in TBG on hBN and to quantify its impact on the electronic spectrum~\cite{long2022atomistic}. The developed theoretical approaches establish the central role of hBN in reshaping the electronic structure of graphene~\cite{moon2014electronic} and twisted bilayer graphene~\cite{long2022atomistic,long2023electronic}, including gap openings at the Dirac point and the appearance of secondary Dirac cones~\cite{Song2013Electron,Amet2013Insulating,Hunt2013Massive,Getal14Detecting,Chen2014Observation,Yankowitz2014Graphene,Wang2016Gaps,Jung2015Origin,Lee2016Ballistic,Kim2018Accurate}.

\section{Computational methods with TB for moir\'e superlattices}
The TB model is a powerful tool for analyzing the physics arising from the moir\'e systems. In particular, the single-particle band structure of the TB Hamiltonian is a good and accurate starting point to describe the moir\'e structure and explain the experimental results. However, in these large-scale and complex systems, the loss of angstrom-scale periodicity and possession of moir\'e-scale period imply that the moir\'e unit cell contains a large number of atoms. Such large-scale TB Hamiltonian matrix poses a significant theoretical challenge. In the following, we review several methods for dealing with these large-size Hamiltonian matrices. 

\subsection{Diagonalization method}

To analyze electronic properties such as the band structures in Fig. \ref{fig:S-p TB}, a typical computational method is directly diagonalizing the full TB Hamiltonian $H_{tb}$ to obtain its eigenvalues $E$ and eigenstates $\psi$ satisfying
\begin{equation}
    H_{tb}\psi = E\psi.
\end{equation}
For the orthogonal basis, this is a dense Hermitian eigen-problem, with the cost of time and memory scaling as $\mathcal{O}(N^3)$ and $\mathcal{O}(N^2)$, respectively. The non-orthogonal TB Hamiltonian leads to a generalized form $H\psi=ES\psi$ with an overlap matrix $S$\cite{golub2013matrix}. When only a small number of eigenpairs near the Fermi level are required, e.g., bands in a narrow energy window or low-frequency transport/optics, \emph{partial-spectrum} solvers are markedly efficient tools for sparse TB Hamiltonian matrices. The Krylov method can target extremal or interior eigenvalues. With a \emph{shift–invert} one iterates on the operator
\begin{equation}
  \left(H-\sigma S\right)^{-1}S ,
\end{equation}
so that eigenvalues closest to the shift $\sigma\!\approx\!E_F$ converge first \cite{lanczos1950iteration,davidson1975iterative,sleijpen2000jacobi,lehoucq1998arpack,stathopoulos2010primme,hernandez2005slepc}. In practice, full diagonalization remains simple and robust for moderate $N$, while partial-spectrum solvers become attractive for very large supercells or dense $k$-meshes focused on a small energy window around $E_F$.

 Once $\{E,\psi\}$ are available, numerous static and dynamical observables can be evaluated via Kubo formulas in the eigenstate basis\cite{kubo1957statistical}. For example, the optical conductivity can be formulated as\cite{greenwood1958boltzmann,kuang2024optical}
\begin{eqnarray}
\sigma _{\alpha1 \alpha2}\left( \omega \right) =\frac{g_si}{ (2\pi )^D}\int_{BZ}{d^D}\mathbf{k}\sum_{l,l'}{\frac{n_F\left( E_{\mathbf{k}l'} \right) -n_F\left( E_{\mathbf{k}l} \right)}{E_{\mathbf{k}l}-E_{\mathbf{k}l'}}}\nonumber\\
\times{\frac{ \left< \mathbf{k}l'\left| J_{\alpha1} \right|\mathbf{k}l \right> \left< \mathbf{k}l\left| J_{\alpha2} \right|\mathbf{k}l' \right> }{E_{\mathbf{k}l'}-E_{\mathbf{k}l}+\hbar \omega +i\delta}}, 
\label{eq:dia_optic}
\end{eqnarray}
where $g_s$ is the spin degeneracy and $D$ is the dimension of {\mo} structure, typically set to 2 for 2D materials. 
$J_{\alpha1}$ and $J_{\alpha2}$ are current operators along the $\alpha1$ and $\alpha2$ directions, respectively. $n_F$ is the Fermi-Dirac distribution. Eigenvalues $E_{\mathbf{k}l}$ and eigenstates $|\mathbf{k}l \rangle$,  with band index $l$ and momentum $\mathbf{k}$, are needed to describe optical band transitions between $l$ and $l'$ bands. The integration runs over the whole Brillouin zone (BZ).

\begin{figure}
     \centering
     \includegraphics[width=0.5\textwidth]{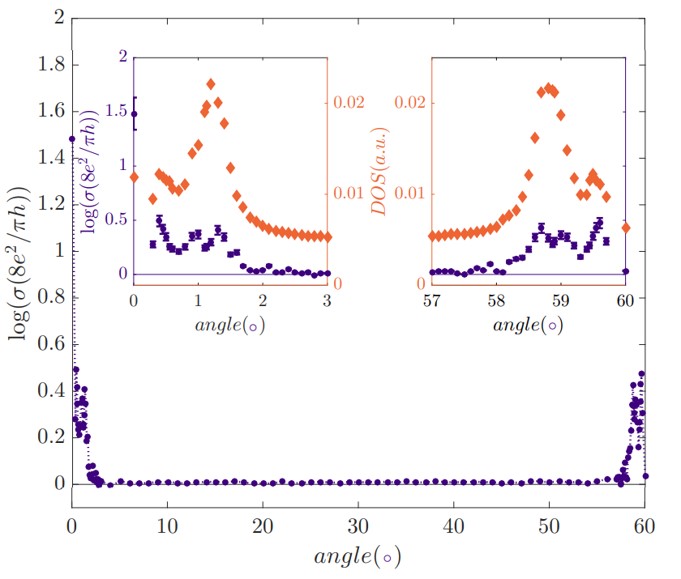}
     \caption{DC conductivity and DOS calculated from KPM based on the TB model introduced in section 2.2.1 for TBG over a wide range of angles. Left and right insets display the DC and DOS for small and large angles, respectively. Adapted with permission from \cite{andjelkovic2018dc}. Copyright (2018) by the American Physical Society.}
     \label{figure:KPM}
\end{figure}

\subsection{Linear-scaling random state methods}
A full diagonalization method will not be very efficient when a moir\'e supercell contains more than thousands of atoms. For example, the number of atoms in TBG increases rapidly when reducing the angle $\theta$. For instance, the angle $\theta \approx 0.22^{\circ}$ contains more than 260,000 atoms. The calculation of electronic structures of TBG with tiny angles is numerically challenging. In this case, a linear-scaling method with scale of $O(N)$ has the advantage of tackling the large-scale TB Hamiltonian \cite{weisse2006kernel,yuan2010modeling,li2023tbplas}.

One of the linear-scaling methods is the random state kernel polynomial method (KPM)\cite{weisse2006kernel}. For example, the DOS can be expressed as
\begin{equation}
D(E)
= \frac{1}{\pi \,\sqrt{1 - E^{2}}}
  \biggl[
    \gamma_{0}^{M}\,\mu_{0}
    \;+\;
    2 \sum_{m=1}^{M}
      \gamma_{m}^{M}\,\mu_{m}\,T_{m}(E)
  \biggr],\label{eq:dos_kernel}
\end{equation}
where $E$ is rescaled to $[-1,1]$ and $\gamma_{m}^{M}$ is a kernel coefficient; a Jackson kernel, widely used, has the form
\begin{equation}
\gamma_{m}^{M}
= \frac{
  (M - m + 1)\,\cos\!\Bigl(\frac{\pi m}{M+1}\Bigr)
  \;+\;
  \sin\!\Bigl(\frac{\pi m}{M+1}\Bigr)
  \,\cot\!\Bigl(\frac{\pi}{M+1}\Bigr)
}{\,M + 1\,},
\end{equation}
where $T_m(E)$ is the Chebyshev polynomial with the recursive relation
\begin{equation}
T_{m}(x) \;=\; 2\,x\,T_{m-1}(x)\;-\;T_{m-2}(x).
\label{eq:Tm}
\end{equation}
Here $T_{m}(x) \;=\; \cos\bigl[m\,\arccos(x)\bigr]$, resulting in $T_{0}(x) = 1$ and $T_{1}(x) = x$. The parameter $\mu_{m}$ is the Chebyshev moment computed through
\begin{equation}
   \mu_{m} = \mathrm{Tr}\bigl[T_{m}(\tilde{H})\bigr] \approx \frac{1}{R} \sum_{p=1}^{R} \langle \psi_p(\boldsymbol{r}) \mid T_{m}(\tilde{H}) \mid \psi_p(\boldsymbol{r}) \rangle, \label{eq:trace_moment}
\end{equation}
where $\psi_p(\boldsymbol{r})$ is the random (stochastic) state of the expanded moir\'e superlattice, and $\tilde{H}$ is a rescaled Hamiltonian with eigenvalues ranging for -1 to 1. The error of this approximation is $\mathcal{O}\bigl(1/\sqrt{RN}\bigr)$, with $R$ the number of random states and $N$ the size of the Hamiltonian. The large-scale moir\'e superlattices naturally give a large $N$ Hamiltonian that benefits the trace of Eq. (\ref{eq:trace_moment}) convergence, but is hard to be diagonalized. A Kubo-Bastin DC conductivity of large-scale moir\'e can be computed with\cite{bastin1971quantum,garcia2015real,andjelkovic2018dc}
\begin{equation}
	\sigma_{\alpha1\alpha2}(\mu,T)=\frac{4e^2\hbar}{\pi A}\frac{4}{\Delta E^2}\int_{-1}^1 \mathrm{d}\tilde{E}\frac{n_F(\tilde E)}{(1- \tilde E^2)^2}\displaystyle\sum_{m,n}\Gamma_{nm}(\tilde E)\mu_{nm}^{\alpha1\alpha2}(\tilde H)
    \label{DC_kernel}
\end{equation}
where $\Delta E=E^+_{max}-E^-_{min}$ is the energy range of the spectrum and $\tilde E$ is the rescaled energy within [-1,1]. $\Gamma_{nm}(\tilde E)$ and $\mu_{nm}^{\alpha1\alpha2}(\tilde H)$ are functions of the energy and the Hamiltonian, respectively
\begin{align}
	\Gamma_{nm}(\tilde E)&=T_m(\tilde E)(\tilde E-\mathrm{i}n\sqrt{1-\tilde E^2})\mathrm{e}^{\mathrm{i}n\arccos{(\tilde E)}} \nonumber\\
	&\;+T_n(\tilde E)(\tilde E+\mathrm{i}m\sqrt{1-\tilde E^2})\mathrm{e}^{-\mathrm{i}m\arccos{(\tilde E)}}, \nonumber\\
	\mu_{nm}^{\alpha_1\alpha_2}(\tilde H)&=\frac{g_mg_n}{(1+\delta_{n0})(1+\delta_{m0})}\mathrm{Tr}[v_{\alpha1} T_m(\tilde H)v_{\alpha2} T_n(\tilde H)],
\end{align}
where $g_m$ can be represented as a Lanczos kernel with
$g_{n} = \frac{\sinh\bigl[\lambda\bigl(1 - \tfrac{n}{N}\bigr)\bigr]}{\sinh(\lambda)}$ and \(\lambda = 4\). $v_{\alpha1}$ is the $\alpha1$ component of velocity operator \(\displaystyle v = -\tfrac{i}{\hbar}\,[\mathbf{l},\,H]\), where $\mathbf{l}$ is the distance vector, and the trace can be calculated in a random state basis through Eq. (\ref{eq:trace_moment})\cite{andjelkovic2018dc}. As shown in Fig. \ref{figure:KPM}, the KPM is a powerful method for modeling the DOS, Direct current (DC) conductivity and conductance in graphene-based moir\'e systems with tiny angles\cite{de2024fast,ho2023optical,le2019real,andjelković2020double,de2023efficient} and could facilitate the computation of properties of more complex mori\'{e} superlattices in the future.

\begin{figure}
     \centering
     \includegraphics[width=0.5\textwidth]{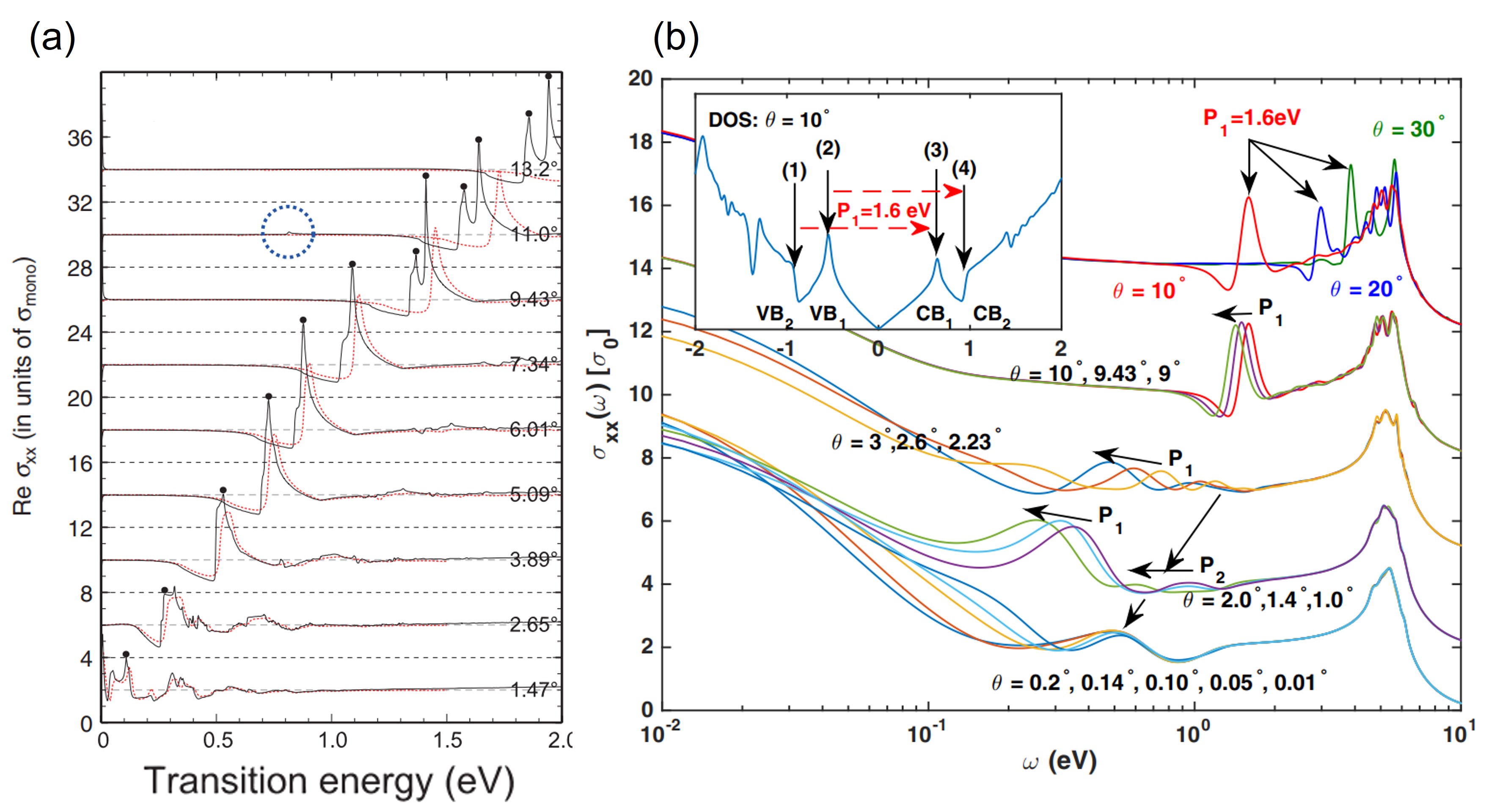}
     \caption{(a) Evolution of the optical conductivity (solid black lines) of TBG with commensurate angles, calculated by using Eq. (\ref{eq:dia_optic}) with exact diagonalization of the TB Hamiltonian. The dashed red circles are the continuum results. Adapted with permission from \cite{moon2013optical}. Copyright (2013) by the American Physical Society.  (b) Evolution of the optical conductivity of TBG with varied angles, calculated by using Eq. (\ref{eq:tbpm_optic}) with a combination of TBPM and TB model. Conductivity peaks corresponding optical transitions between VHS of DOS are indicated by arrows in the inset. Adapted with permission from \cite{le2018electronic}. Copyright (2018) by the American Physical Society.}
     \label{figure:tbpm}
\end{figure}

The tight-binding propagation method (TBPM) is another powerful approach to simulate the broad properties of large-scale moir\'e materials\cite{li2023tbplas}. Compared to KPM, a time-evolution is applied to extract the information of a simulated system\cite{yuan2010modeling}. For example, the DOS can be calculated as \cite{yuan2010modeling,li2023tbplas}
\begin{equation}
	D(E) = \frac{1}{R} \sum_{p=1}^{R} \frac{1}{2\pi} \int_{-\infty}^{\infty} \mathrm{e}^{\mathrm{i}Et} \langle \varphi_p(0)\mid e^{-\,i\,H\,t}\mid \varphi_p(0)\rangle \mathrm{d}t,
    \label{eq:dos_tbpm}
\end{equation}
where $\varphi_p(0)$ is the $p$th initial random state at $t=0$. The calculation converges with an increasing number of random samples $R$ and the size of the Hamiltonian $N$. Based on TBPM, the optical conductivity can be calculated\cite{yuan2010modeling,li2023tbplas}
\begin{eqnarray}
	\sigma_{\alpha1 \alpha2}\left( \omega \right) =\underset{\varepsilon \rightarrow 0^+}{lim}\frac{e^{-\beta \hbar \omega}-1}{\hbar \omega \Omega}\int_0^{\infty}{e^{-\varepsilon t}}\left( \sin\omega t-i\cos\omega t \right) \nonumber \\
	\times 2\mathrm{Im}\left\{\langle \varphi \left| n_F\left( H \right) e^{iHt}J_{\alpha1}e^{-iHt}[1-n_F\left( H \right) ]J_{\alpha2} \right|\varphi \rangle\right\} dt. \label{eq:tbpm_optic}
\end{eqnarray}
Here, $\Omega$ is the area or volume of the model, $\beta=1/k_BT$ with $k_B$ the Boltzmann constant and $T$ the temperature. Compared to the $O(N^3)$ time scaling of Eq. (\ref{eq:dia_optic}), the random-state method scales linearly $O(N)$ with the dimension of Hamiltonian in real space. Besides, TBPM can be applied to calculate dynamical properties in both commensurate and incommensurate moir\'{e} supperlatice (see Fig. \ref{figure:tbpm}(b)), while the diagonalization method in reciprocal space can only work for commensurate ones (see Fig. \ref{figure:tbpm}(a)). The merits and flexibility of TBPM also make it validly explain experimental phenomena and simulate broad electronic and dynamical properties in various moir\'e materials (see section 5)\cite{shi2020large,yu2020electronic,hu2023observation,cui2024nanoscale,hao2024robust}.

\subsection{Tight-binding methods with machine learning}
\begin{figure}
     \centering
     \includegraphics[width=0.5\textwidth]{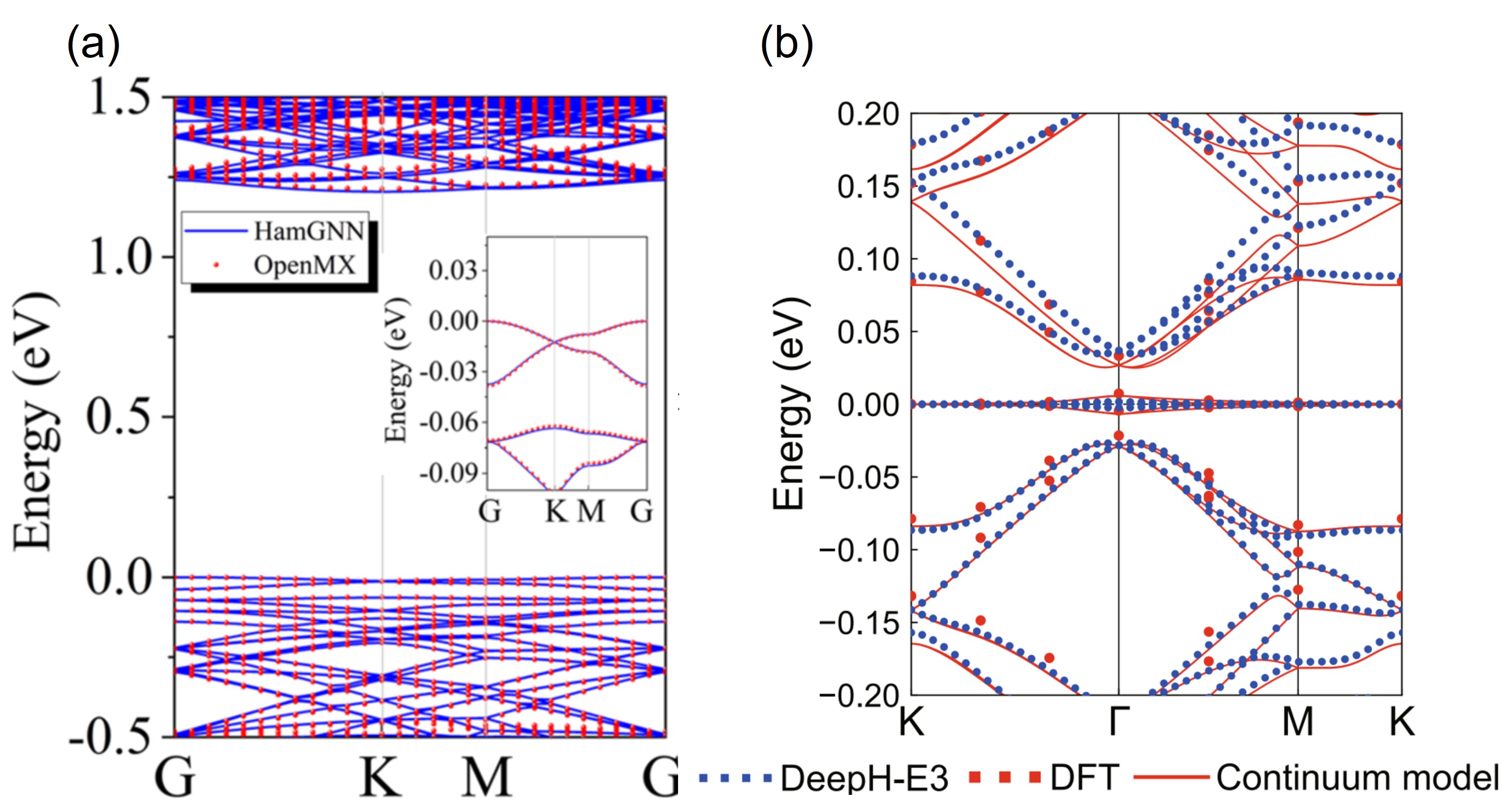}
     \caption{(a) Comparison of band structures of twised bilayer MoS$_2$ at $\theta=3.5^\circ$ obtained from machine-learning HamGNN method (lines) and DFT calculation (dots). A band zoom near zero is shown in the inset. Adapted under the terms of the CC BY license from \cite{zhong2023transferable}. Copyright (2023) the authors. (b) Bands of TBG at $\theta=1.08^\circ$ predicted by DeepH method, compared to those obtained from DFT calculation (red dots) and continuum model (red lines) (Details of continumm model in next section). Adapted under the terms of the CC BY license from \cite{gong2023general}. Copyright (2023) the authors.}
     \label{figure:MLTB}
\end{figure}

A convincing atomic TB model is relevant for exploring properties of mori\'{e} supperlatices. Recently, machine learning methods have emerged to favor the construction of TB Hamiltonian and investigate the electronic properties of moir\'e superlattices\cite{yang2024evolution,liu2022seeing,tritsaris2020electronic,li2022deep,li2023deep,tang2024deep,gobbo2025layer}. For example, by training various small bilayer stackings of graphene, deep learning-based methods such as DeepH can reproduce electronic structures of a large-scale TBG moir\'{e} up to DFT accuracy (see Fig. \ref{figure:MLTB}(b))\cite{li2022deep,li2023deep,gong2023general}. By similarly preparing the train dataset from real-space DFT calculation as DeepH, HamGNN method can also train and infer the \textit{ab initio} accuracy TB Hamiltonian of large-scale moir\'e materials such as twisted bilayer MoS$_2$ as displayed in Fig. \ref{figure:MLTB}(a) \cite{zhong2023transferable,zhong2024universal}. While the so-called ab-initial TB Hamiltonian from DeepH and HamGNN is actually a numerical TB Hamiltonian expanded in a group of non-orthogonal and overlapped localized basis, a DeepTB method can generate a semi-empirical SK TB Hamlitonian with ab initio accuracy over a wide range of elements, which could open new possibility to provide accurate SK parameters for generating Hamiltonian for unknown moir\'e materials\cite{gu2024deep}.

\subsection{Software packages within TB for modeling moire superlattices}
Atomically modeling a moir\'e material based on TB Hamiltonian contains some typical tasks including the construction of the superlattice, relaxation, building a TB Hamiltonian, employing numerical methods to study properties and postprocessing. There are some useful and versatile software packages facilitating these modeling tasks. $Twister$ is specialized to construct and relax a moir\'e superlattice\cite{naik2022twister}. Recently, $DPmoire$ provides a means to generate ab-initial accuracy machine-learning force fields specifically tailored for moir\'e structures\cite{liu2024dpmoire}, which interfaces with molecular dynamics software such as $Lammps$\cite{thompson2022lammps} and $ASE$ \cite{larsen2017ase} for atomic relaxation. The versatile package $KITE$ incorporates the atomic construction of a moir\'e superlattice, KPM for calculating transport and optical properties and visualization\cite{joao2020kite}. It also provides the interface with other packages such as $Pybinding$, which is also based on TB methods with both the exact diagonalization and the KPM\cite{jakob2024pybind11}. $TBPLaS$ is a functional package covering all the procedures required to simulate a moir\'e superlattice\cite{li2023tbplas}. It features with exact diagonalization, TBPM, and KPM to calculate various properties of moir\'e superlattice. It also has interface with $Wannier90$\cite{pizzi2020wannier90}, $Lammps$\cite{thompson2022lammps}, $DeepH$\cite{li2022deep} and $DeepTB$\cite{gu2024deep} to keep its flexibility in considering relaxation and constructing a new Hamiltonian for a moir\'e superlattice.\\

\section{Fitting TB to low-energy continuum models}

The interesting regime of low twist angles in moiré superlattices leads to very large moiré lengths, with up to thousands of atoms per supercell. This naturally imposes a heavy computational cost on atomistic TB simulations. Besides time-consuming limitations, dealing with huge supercells can hinder an intuitive understanding on how the system behavior changes as the twist angle decreases. In addition, going beyond the TB single-particle picture becomes exponentially more difficult as the number of atoms increase. Yet, it is at these large moiré superlattices where the electronic correlations become crucial.

\begin{figure}[t]
    \centering
        \includegraphics[width=\linewidth]{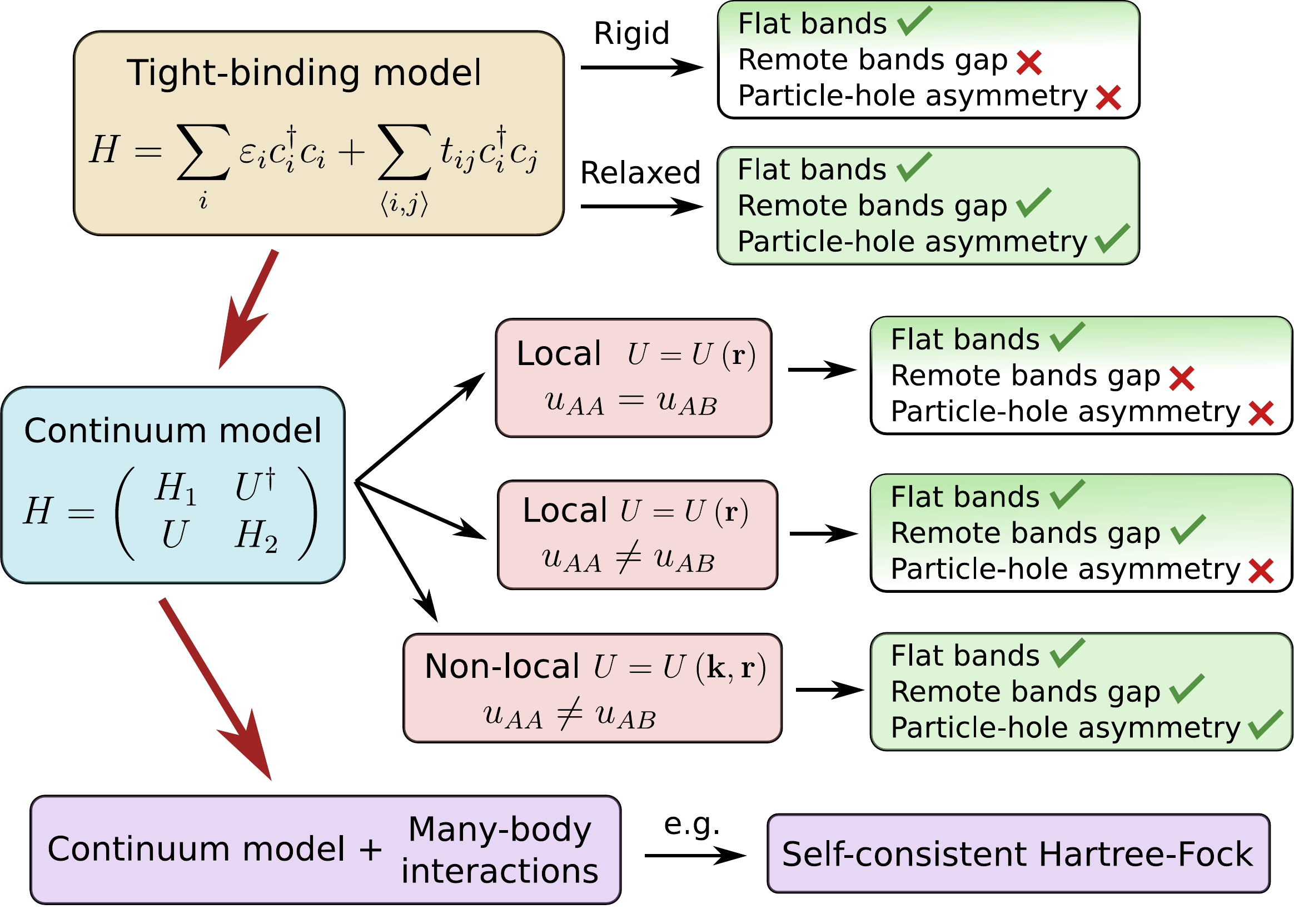}
    \caption{Schematic representation of the path from TB models to effective continuum models in TBG. Different approximations are gauged by the properties of the band structure around the magic angle: the emergence of flat bands, their gap with the remote bands, and the particle-hole asymmetry. The later two properties only emerge in the TB model when the system is allowed to relax. The continuum model provides a low-energy description in which the two layers, with Dirac Hamiltonians $H_{1}$ and $H_{2}$, are coupled by a moiré potential $U$ with effective hoppings $u_{AA}$ and $u_{AB}$ at the AA and AB/BA stacking regimes. Capturing the three main properties of the flat bands depends, primarily, on the ratio between the hopping energies and the locality of the moiré potential \cite{koshino2020effective, garcia2021full, zhu2024weak}. From the simple continuum model one can then more easily go beyond the single-particle picture by taking into account many-body interactions.}\label{fig:TB_to_CM}
\end{figure}

These considerations have motivated the need of having effective continuum descriptions of the electronic properties in moiré systems, which can capture the TB results, but yet are simpler enough to allow efficient extensions of it by including, for instance, correlations effects. Having simpler continuum models can also provide valuable insights on the nature and origin of flat bands in moiré systems\cite{tarnopolsky2019origin, wang2021chiral, becker2021spectral, watson2021existence, wang2024role, escudero2024diagrammatic}. Furthermore, a continuum model can be constructed even if the systems is incommensurate\cite{bistritzer2011moire, koshino2015interlayer}. A simple schematic hierarchy of the fitting of TB models to low-energy continuum models is shown in Figure \ref{fig:TB_to_CM}. 

The continuum description rest upon the fact at low twist angles the moiré scale becomes much larger than the atomic length, so the interlayer interaction is dominated by its long wavelength components\cite{bistritzer2011moire, koshino2015interlayer}. This means that the electronic behavior can be well described by the continuum approximation. The continuum description was originally introduced for TBG in 2007 by Lopes dos Santos \textit{et al.}\cite{lopes2007graphene}, for commensurate structures, and later extended to account for incommensurate structures by Bistritzer and MacDonald in 2011\cite{bistritzer2011moire}. The later model allows one to define a moiré Brillouin zone and obtain the band structure of TBG for \emph{any} low twist angle. These pioneering formulations not only captured the low energy spectra obtained by the TB model, but also allowed one to obtain further simpler models of the flat bands as linear dispersions with a renormalized velocity that vanishes at the magic angle $\theta\sim1.05^{\circ}$\cite{lopes2007graphene, bistritzer2011moire}. Since then, many other works have reformulated \cite{lopes2012continuum, moon2013optical, koshino2018maximally, kang2018symmetry, carr2019exact, garcia2021full, bernevig2021twisted, miao2023truncated, zhu2024weak}, and extended these continuum models to account for large twist angles\cite{koshino2015interlayer}, lattice relaxation\cite{guinea2019continuum, fang2019angle, koshino2020effective, kang2023pseudomagnetic, ceferino2024pseudomagnetic}, and strain effects\cite{bi2019designing, balents2019general, pantaleon2021tunable, mannai2021twistronics, mesple2021heterostrain, sinner2023strain, vafek2023continuum, escudero2024designing}. The original continuum model of TBG has been further extended to other moiré structures, such as twisted TMDs\cite{wu2019topological, devakul2021magic, angeli2021gamma}, twisted hBN\cite{walet2021flat}, twisted graphene/hBN\cite{jung2017moire, cea2020band, long2022atomistic, long2023electronic}, and twisted multilayer graphene \cite{waters2024topological, su2025moire}. In what follows we focus on TBG and mostly follow the continuum formulation of Koshino \textit{et al.}\cite{koshino2018maximally}.

The starting point is to define the Bloch wave states in each layer as
\begin{equation}
\left|\mathbf{k},X\right\rangle =\frac{1}{\sqrt{N}}\sum_{\mathbf{R}_{X}}e^{i\mathbf{k}\cdot\mathbf{R}_{X}}\left|\mathbf{R}_{X}\right\rangle ,
\end{equation}
where $X=\left\{ A_{1},B_{1},A_{2},B_{2}\right\} $ is the layer-sublattice index, $N$ is the number of graphene monolayer cells in each layer, and $\mathbf{R}_{l}$ are the atomic positions
\begin{align}
\mathbf{R}_{A_{1}} & =n_{1}\mathbf{a}_{1}+n_{2}\mathbf{a}_{2}+\boldsymbol{\tau}_{A_{1}},\nonumber \\
\mathbf{R}_{B_{1}} & =n_{1}\mathbf{a}_{1}+n_{2}\mathbf{a}_{2}+\boldsymbol{\tau}_{B_{1}},\nonumber \\
\mathbf{R}_{A_{2}} & =n_{1}\mathbf{a}_{1}+n_{2}\mathbf{a}_{2}+\boldsymbol{\tau}_{A_{2}}+\boldsymbol{\delta}+d\left(\boldsymbol{\delta}\right)\mathbf{e}_{z},\nonumber \\
\mathbf{R}_{B_{2}} & =n_{1}\mathbf{a}_{1}+n_{2}\mathbf{a}_{2}+\boldsymbol{\tau}_{B_{2}}+\boldsymbol{\delta}+d\left(\boldsymbol{\delta}\right)\mathbf{e}_{z},
\end{align}
where $\mathbf{a}_{1}=a\left(1,0\right)$ and $\mathbf{a}_{2}=a\left(1/2,\sqrt{3}/2\right)$ are the monolayer's lattice vectors, while $\boldsymbol{\tau}_{X}$ are the sublattice displacements ($\boldsymbol{\tau}_{A_{1}}=\boldsymbol{\tau}_{A_{2}}=0$, $\boldsymbol{\tau}_{B_{1}}=\boldsymbol{\tau}_{B_{2}}=-\boldsymbol{\tau}_{1}$ with $\boldsymbol{\tau}_{1}=\left(2\mathbf{a}_{2}-\mathbf{a}_{1}\right)/3$). The displacement vector $\boldsymbol{\delta}$ accounts for the variation in the atomic positions of layer 2 due to its relative rotation with layer 1, while $d\left(\boldsymbol{\delta}\right)$ accounts for the interlayer distance at $\boldsymbol{\delta}$. When the layers are relatively rotated by a small twist angle $\theta$, the displacement vector $\boldsymbol{\delta}$ is taken to vary with the real space position $\mathbf{r}$ as\cite{koshino2018maximally}
\begin{equation}
\boldsymbol{\delta}\left(\mathbf{r}\right)=\left[R\left(\theta/2\right)-R\left(-\theta/2\right)\right]\mathbf{r}.
\end{equation}
Due to relaxation effects, the corresponding interlayer distance $d\left(\boldsymbol{\delta}\right)$ is not uniform throughout the supercell: it is maximum around the AA stacking with $d_{AA}=0.36\,\text{nm}$, and minimum around the AB stacking $d_{AB}=0.335\,\text{nm}$. Koshino \textit{et al.}\cite{koshino2018maximally} interpolated $d$ as 
\begin{equation}
d\left(\boldsymbol{\delta}\right)=d_{0}+2d_{1}\sum_{j=1}^{3}\cos\left(\mathbf{b}_{i}\cdot\boldsymbol{\delta}\right),
\end{equation}
where $\mathbf{b}_{3}=-\mathbf{b}_{1}-\mathbf{b}_{2}$, $d_{0}=\left(d_{AA}+2d_{AB}\right)/3$ and $d_{1}=\left(d_{AA}-d_{AB}\right)/9$. 

\begin{figure}[t]
    \centering
        \includegraphics[width=0.9\linewidth]{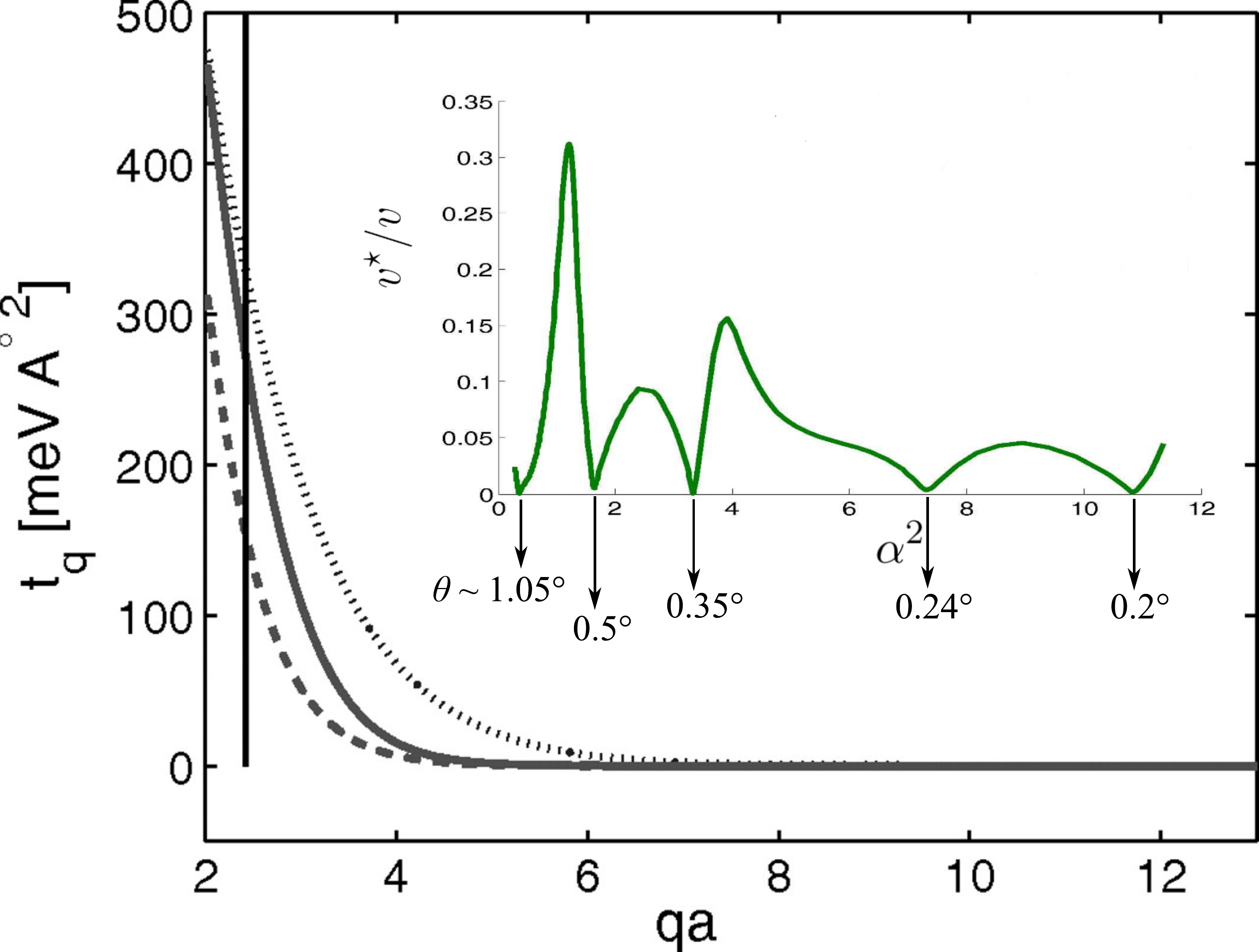}
    \caption{Dependence of the moiré-induced interlayer tunneling on the momentum $qa=\left|\mathbf{q}\right|a$, where $a\simeq0.142\,\mathrm{nm}$ is the carbon-carbon distance in graphene. The solid, dashed and dot lines correspond to the models described in Refs. \cite{pereira2009tight}, \cite{tang1996environment} and \cite{bistritzer2010transport}, respectively. The vertical lines indicates the point $k_{D}a$, where $k_{D}=\left|\mathbf{K}\right|$ is the distance of the monlayer's Dirac point. Inset shows the renormalized Fermi velocity $v^{\star}$ obtained by the Bistritzer-MacDonald continuum model, predicting a series of magic angles where $v^{\star}$ vanishes. Adapted under the terms of the CC BY license from \cite{bistritzer2011moire}. Copyright (2011) National Academy of Sciences.} \label{fig:tq_BMmodel}
\end{figure}

Assuming that the transfer integral between sites $\mathbf{R}_{X}$ and $\mathbf{R}_{X'}$ depends only on their relative distance, the interlayer matrix elements that couple the two layers takes the form\cite{moon2013optical, koshino2015interlayer}
\begin{equation}
U=-\sum_{X,X'}t\left(\mathbf{R}_{X'}-\mathbf{R}_{X}\right)\left|\mathbf{R}_{X'}\right\rangle \left\langle \mathbf{R}_{X}\right|+h.c.,
\end{equation}
where the transfer integral $t\left(\mathbf{R}\right)$ is given by the SK parametrization in the TB model. Replacing the plane-wave expansion of the Bloch states $\left|\mathbf{R}_{X}\right\rangle $, and using the continuum description of the displacement vector $\boldsymbol{\delta}\left(\mathbf{r}\right),$ leads to the interlayer interaction
\begin{align}
U_{X'X}\left(\mathbf{k}',\mathbf{k}\right) & \equiv\left\langle \mathbf{k}',X'\right|U\left|\mathbf{k},X\right\rangle \nonumber \\
 & =\sum_{m_{1},m_{2}}t_{X'X}\left(\mathbf{k}+m_{1}\mathbf{b}_{1}+m_{2}\mathbf{b}_{2}\right)\nonumber \\
 & \times e^{i\left(m_{1}\mathbf{b}_{1}+m_{2}\mathbf{b}_{2}\right)\cdot\left(\boldsymbol{\tau}_{X'}-\boldsymbol{\tau}_{X}\right)}\delta_{\mathbf{k}'-\mathbf{k},m_{1}\mathbf{G}_{1}+m_{2}\mathbf{G}_{2}},\label{eq:moireP_g}
\end{align}
where $\mathbf{G}_{i}=\left[R\left(\theta/2\right)-R\left(-\theta/2\right)\right]\mathbf{b}_{i}$ are the moiré vectors and $t_{X'X}\left(\mathbf{q}\right)$ is the in-plane Fourier transform of the transfer integral
\begin{equation}
t_{X'X}\left(\mathbf{q}\right)=-\frac{1}{S_{0}}\int d\mathbf{r}t\left[\mathbf{r}+d\left(\mathbf{r}-\boldsymbol{\tau}_{X'}+\boldsymbol{\tau}_{X}\right)\right]e^{-i\mathbf{q}\cdot\mathbf{r}},
\end{equation}
where $S_{0}=\left(\sqrt{3}/2\right)a^{2}$ in the unit cell of monolayer graphene. Figure \ref{fig:tq_BMmodel} shows the variation of the hopping amplitude $t\left(\mathbf{q}\right)$ as a function of momentum $q=\left|\mathbf{q}\right|$, for different models. The key observation is that $t\left(\mathbf{q}\right)$ decays very rapidly with $q$ because the interlayer separation exceeds the intralayer carbon-carbon distance by more than a factor of 2\cite{bistritzer2011moire}. 

\begin{figure*}[t]
    \centering
        \includegraphics[width=\linewidth]{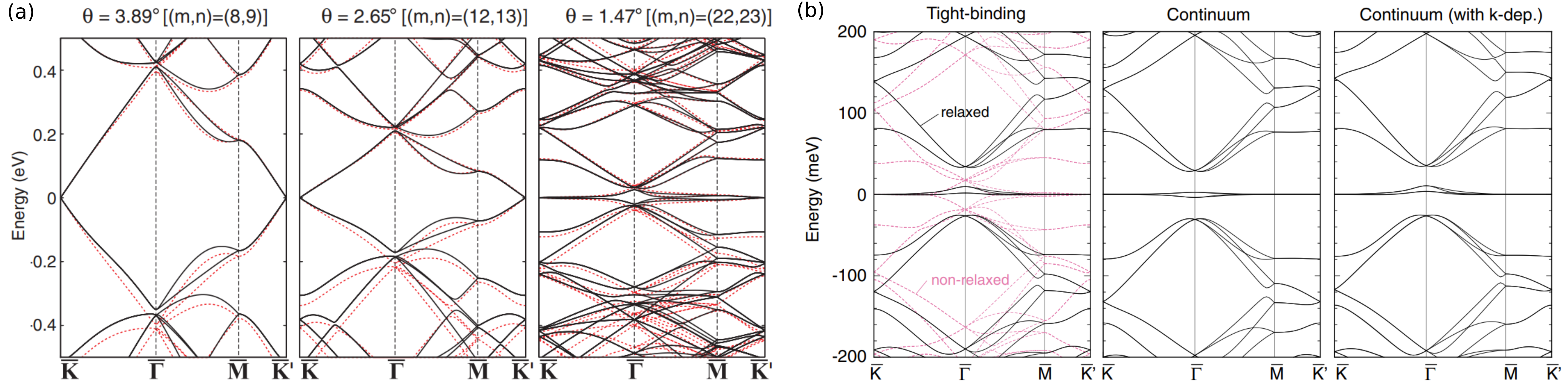}
    \caption{(a) Band structures of rigid twisted bilayer graphene for different commensurate angles $\theta$. The black solid line correspond to the tight-binding results, while the red dotted-line corresponds to the continuum model results with a local moiré potential. Adapted from Ref. \cite{moon2013optical}. (b) Comparison between the band structures of rigid and relaxed twisted bilayer graphene at the magic angle $\theta=1.05^{\circ}$, obtained by the tight-binding model, and the continuum model with local and non-local ($k-$dependent) moiré potential. Only the later captures the relaxed particle-hole asymmetry of the tight-binding flat bands. Adapted from Ref. \cite{koshino2020effective}.}\label{fig:TB_bands_CM}
\end{figure*}

Following the Dirac approximation, the momenta in both layers is measured with respect to their Dirac points $\mathbf{K}_{\xi}$ (where $\xi$ is the valley index), and the transfer integral in Eq. \eqref{eq:moireP_g} is approximated as $\sim t_{X'X}\left(\mathbf{K}_{\xi}+m_{1}\mathbf{b}_{1}+m_{2}\mathbf{b}_{2}\right)$, leading to a local moiré potential
\begin{align}
U_{X'X}\left(\mathbf{r},\xi\right) & =\sum_{m_{1},m_{2}}t_{X'X}\left(\mathbf{K}_{\xi}+m_{1}\mathbf{b}_{1}+m_{2}\mathbf{b}_{2}\right)\nonumber \\
 & \times e^{i\left(m_{1}\mathbf{b}_{1}+m_{2}\mathbf{b}_{2}\right)\cdot\left(\boldsymbol{\tau}_{X'}-\boldsymbol{\tau}_{X}\right)}e^{i\left(m_{1}\mathbf{G}_{1}+m_{2}\mathbf{G}_{2}\right)\cdot\mathbf{r}}.\label{eq:moireP}
\end{align}
The coupling amplitude $t_{X'X}\left(\mathbf{K}_{\xi}+m_{1}\mathbf{b}_{1}+m_{2}\mathbf{b}_{2}\right)$  only depends on the distance of the Dirac points to the origin. As $t_{X'X}\left(\mathbf{q}\right)$ decay rapidly with $q$, one can take only the first three leading terms $\left(m_{1},m_{2}\right)=\left\{ \left(0,0\right),\xi\left(1,0\right),\xi\left(1,1\right)\right\} $ in the summation over $m_{1}$ and $m_{2}$. The moiré potential, in matrix form, then takes the well-known form\cite{koshino2015interlayer, koshino2018maximally}
\begin{equation}
U\left(\mathbf{r},\xi\right)=U_{0}+U_{1}e^{i\xi\mathbf{G}_{1}\cdot\mathbf{r}}+U_{2}e^{i\xi\left(\mathbf{G}_{1}+\mathbf{G}_{2}\right)\cdot\mathbf{r}},
\end{equation}
where\cite{lopes2007graphene, bistritzer2011moire}
\begin{equation}
U_{j}=\left(\begin{array}{cc}
u_{0} & u_{1}e^{-i\phi_{j}}\\
u_{1}e^{i\phi_{j}} & u_{0}
\end{array}\right),
\end{equation}
with $\phi_{j}=\left(j-1\right)2\pi/3$, and $u_{0}$ and $u_{1}$ are the AA and AB/BA stacking amplitudes given by\cite{koshino2018maximally}
\begin{align}
u_{0} & =-\frac{1}{S_{0}}\int d\mathbf{r}t\left[\mathbf{r}+d\left(\mathbf{r}\right)\mathbf{e}_{z}\right]e^{-i\mathbf{K}_{\xi}\cdot\mathbf{r}},\\
u_{1} & =-\frac{1}{S_{0}}\int d\mathbf{r}t\left[\mathbf{r}+d\left(\mathbf{r}-\boldsymbol{\tau}_{1}\right)\mathbf{e}_{z}\right]e^{-i\mathbf{K}_{\xi}\cdot\mathbf{r}}.
\end{align}
Koshino \textit{et al.}\cite{koshino2018maximally} obtained $u_{0}=0.0797\,\mathrm{eV}$ and $u_{1}=0.0975\,\mathrm{eV}$. Note that for flat TBG, as considered initially in the Bistritzer-MacDonald model\cite{bistritzer2011moire}, the interlayer distance $d\left(\mathbf{r}\right)$ is constant and thus $u_{0}=u_{1}$.

Finally, the effective continuum model Hamiltonian for the $\xi$ valley takes the form\cite{koshino2018maximally}
\begin{equation}
H_{\xi}=\left(\begin{array}{cc}
H_{1} & U^{\dagger}\\
U & H_{2}
\end{array}\right),
\end{equation}
where $H_{l}$ is the intralayer Dirac Hamiltonian in layer $l=1,2$, given by the two-dimensional Weyl equation centered at the $\mathbf{K}_{l,\xi}$ point
\begin{equation}
H_{l}=-\hbar v\left[R\left(l\theta/2\right)\left(\mathbf{k}-\mathbf{K}_{l,\xi}\right)\right]\cdot\left(\xi\sigma_{x},\sigma_{y}\right).
\end{equation}
Here $\sigma_{x}$ and $\sigma_{y}$ are the Dirac matrices acting on the sublattice space, and\cite{moon2013optical} 
\begin{equation}
v\simeq\frac{\sqrt{3}}{2}\frac{a}{\hbar}V_{pp\pi}^{0}\left(1-2e^{-a_{0}/\delta_{0}}\right)
\end{equation}
is the Fermi velocity, where $a_{0}=a/\sqrt{3}$ is the carbon-carbon distance and $\delta_{0}=0.184a$ is the decay length \cite{moon2013optical}, so that the nearest intralayer coupling is $0.1V_{pp\pi}^{0}$. With $V_{pp\pi}^{0}\sim-2.7\,\mathrm{eV}$, Koshino \textit{et al.} obtained $\hbar v/a=2.1354\,\mathrm{eV}$\cite{koshino2018maximally}. 

To compute the energy bands in the continuum model one expands the Bloch states in plane-waves as
\begin{equation}
\psi_{n\mathbf{k}}^{X}\left(\mathbf{r}\right)=\sum_{\mathbf{G}}C_{n\mathbf{k}}^{X}\left(\mathbf{G}\right)e^{i\left(\mathbf{k}+\mathbf{G}\right)\cdot\mathbf{r}},
\end{equation}
where $n$ is the moiré band index and $\mathbf{k}$ is a momentum vector in the moiré Brillouin zone. Since each state with momentum $\mathbf{k}$ in one layer is coupled, through the moiré potential, to another state with momentum $\mathbf{k}+\mathbf{G}$ in the other layer, the continuum model Hamiltonian in reciprocal space has no inherent cutoff (any state can be always coupled to another through umklapp processes). However, the relevant low-energy spectra is dominated by the coupling of the states closest to the Dirac points, so in practice it is sufficient to consider a large enough momentum cutoff (e.g., $\left|\mathbf{k}\right|<4\left|\mathbf{G}_{1}\right|$), up to which the low-energy spectra converges. The caveat is that the lower the twist angle, the stronger the moiré coupling becomes, and thus the more reciprocal vectors one needs to consider for convergence. This again leads to a high-dimension continuum model Hamiltonian (albeit still much smaller than those in the TB models), further motivating yet simpler effective models for the flat bands \cite{san2013helical, efimkin2018helical, de2021network, bernevig2021twisted, song2022magic, chou2023kondo, hu2023kondo, bennett2024twisted, escudero2024diagrammatic, lau2025topological}.

The moiré potential given by Eq. \eqref{eq:moireP} corresponds to the zeroth order approximation in momenta, i.e., taking $\mathbf{k}\sim\mathbf{K}$ in the general expression given by Eq. \eqref{eq:moireP_g}. As noted, this results in a local, momentum-independent interlayer tunneling. Although this approximation already captures very well the TB spectra (specially the emergence of flat bands around the magic angle; see Figure \ref{fig:TB_bands_CM}), it still cannot capture other important features of the band structure, such as the particle-hole asymmetry of the flat bands due to relaxation effects. To capture such behavior one needs to take into account the contribution of the non-local interlayer tunnelings. 

\begin{figure*}[t]
    \centering
        \includegraphics[width=\linewidth]{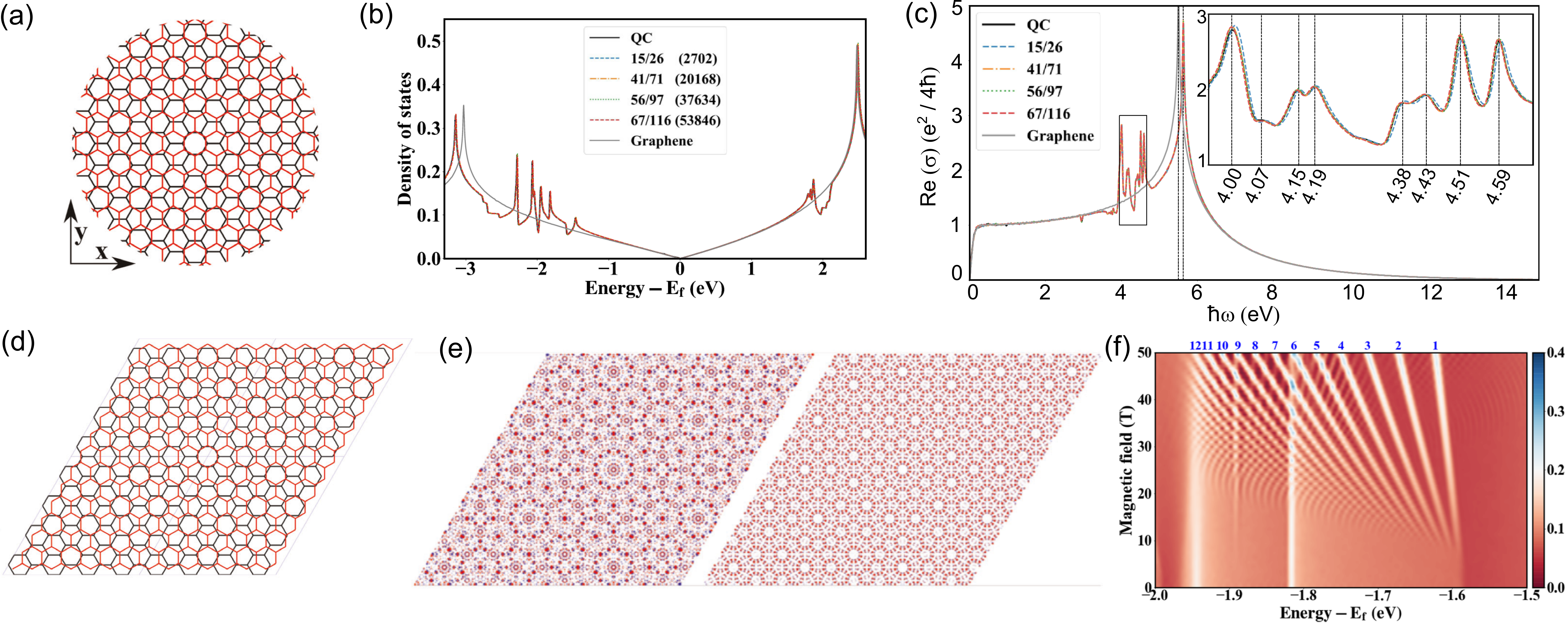}
    \caption{(a) Graphene quasicrystal. (b) DOS obtained from graphene quasicrystal and its approximants. The number of atoms in each unit cell of approximants are in brackets. The DOS of pristine graphene is also plotted. (c) The optical conductivities of graphene quasicrystal, its approximants and graphene. (d) Atomic structure of 4/7 approximant with four unit cells. (e) The eigenstates of 41/71 approximant at -4.2 and -2.76 eV. Red and blue circles represent the states from the top and bottom layers, respectively. (f) Hofstadter's butterflies of 41/71 approximant with magnetic field less than 50 T. Colorbar represents the value of DOS. The blue numbers indicate the indexes of the corresponding Landau levels. Adapted under the terms of the CC BY license from \cite{yu2019dodecagonal}. Copyright (2019) the authors.}
    \label{fig:quasicrystal}
\end{figure*}

The leading order, non-local term follows by expanding the interlayer tunneling $t_{X'X}\left(\mathbf{k}\right)$ around the Dirac point $\mathbf{k}=\mathbf{K}$ up to first order in momenta\cite{fang2019angle, koshino2020effective, garcia2021full, zhu2024weak}
\begin{equation}
t_{X'X}\left(\mathbf{k}\right)\simeq t_{X'X}\left(\mathbf{K}\right)+t'_{X'X}\left(\mathbf{K}\right)\left(k-K\right),
\end{equation}
where
\begin{equation}
t'_{X'X}\left(\mathbf{K}\right)=\left.\frac{\partial t'_{X'X}}{\partial k}\right|_{k=K}<0
\end{equation}
is the non-local tunneling parameter and $K=\left|\mathbf{K}\right|$. Keeping still the three leading-order Fourier components, the momentum-space matrix elements of the moiré potential then become
\begin{align}
U_{X'X}\left(\mathbf{k}',\mathbf{k}\right) & =\sum_{j=1}^{3}\left[t_{X'X}\left(K\right)+t'_{X'X}\left(K\right)\left(\left|\mathbf{k}+\tilde{\mathbf{b}}_{j}\right|-K\right)\right]\nonumber \\
 & \times e^{i\tilde{\mathbf{b}}_{j}\cdot\left(\boldsymbol{\tau}_{X'}-\boldsymbol{\tau}_{X}\right)}\delta_{\mathbf{k}'-\mathbf{k},\mathbf{\tilde{G}}_{j}},
\end{align}
where $\mathbf{\tilde{G}}_{1}=0$, $\mathbf{\tilde{G}}_{2}=\xi\mathbf{G}_{1}$, $\mathbf{\tilde{G}}_{3}=\xi\left(\mathbf{G}_{1}+\mathbf{G}_{2}\right)$ and $\mathbf{\tilde{b}}_{1}=0$, $\mathbf{\tilde{b}}_{2}=\xi\mathbf{b}_{1}$, $\mathbf{\tilde{b}}_{3}=\xi\left(\mathbf{b}_{1}+\mathbf{b}_{2}\right)$. Jihang Zhu \textit{et al.}\cite{zhu2024weak} estimated the non-local tunneling energies as $t'_{AA}g_{M}=-12\,\mathrm{meV}$ and $t'_{AB}g_{M}=-20\,\mathrm{meV}$, where $g_{M}=\left|\mathbf{G}_{1}\right|$. Figure \ref{fig:TB_bands_CM}(b) show the continuum band structure, at the magic angle $\theta=1.05^{\circ}$, with and without the non-local moiré potential; only the non-local potential effectively captures the particle-hole asymmetry obtained in the relaxed TB models.\\

\section{Examples of using TB model in moir\'e systems}
In this section, we provide two examples of using the TB model to study the moir\'e systems. The first example is the theoretical investigation of the electronic properties of graphene quasicrystal\cite{yu2019dodecagonal}, and the second example is the theoretical explanation of the Rydberg moir\'e excitons in WSe$_2$/TBG heterostructure\cite{hu2023observation}. 

\subsection{Dedocagonal bilayer graphene quasicrystal}
When the AA stacking bilayer graphene rotates with an angle of $\theta=30^{\circ}$, a dodecagonal bilayer graphene quasicrystal is formed (see Fig. \ref{fig:quasicrystal}(a)). Interestingly, the dodecagonal graphene quasicrystal has a 12-fold rotational symmetry but lacks translational symmetry. The dodecagonal graphene quasicrystal has been investigated by experiments, showing distinct properties from graphene\cite{yao2018quasicrystalline,ahn2018dirac}. The lack of translational symmetry prevents the application of band theory and requires a new method in this system. In 2019, Yu and coworkers explicitly studied the electronic properties of the dedocagonal graphene quasicrystal\cite{yu2019dodecagonal}. First, by combining the TBPM and TB methods, they studied the electronic and optical properties (Figs. \ref{fig:quasicrystal}(b) and (c)). In particular, to accurately calculate the characteristics, we adopted a large round disk of graphene quasicrystal with ten million atoms described by the TB Hamiltonian. Such large dimension of TB Hamiltonian was solved by the TBPM method. As shown in Fig. \ref{fig:quasicrystal}(b), compared to the graphene case, the graphene quasicrystal possessed distinct peaks in the DOS spectrum around $\pm$ 2 eV, which were attributed to the interlayer interaction. In the vicinity of the Fermi level, the DOS was almost the same as the pristine graphene, which indicated that the optical conductivity at low energies was also the same (see Fig. \ref{fig:quasicrystal}(c)). Importantly, peaks emerged around 4.0 $\approx$ 4.6 eV in the optical spectrum, which were attributed to the VHS of quasicrystal states. 

\begin{figure*}[t]
    \centering
    \includegraphics[width=0.9\linewidth]{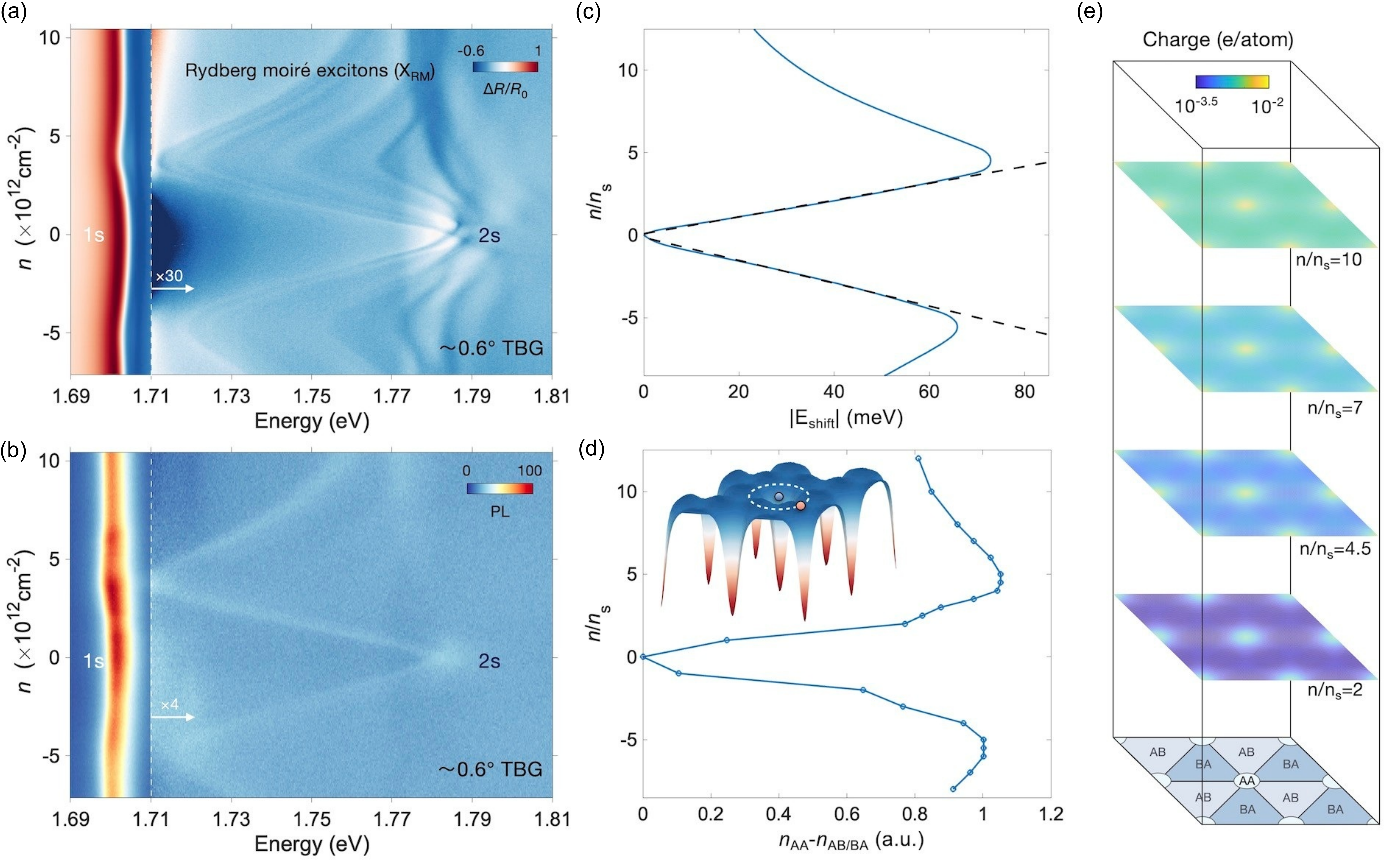}
    \caption{(a) Reflectance contrast spectrum of WSe$_2$/TBG heterostructure with the angle $\theta=0.6^{\circ}$ in TBG. X$_{\mathrm{RM}}$ is the spatial confinement of Rydberg moir\'e excitons. (b) Photoluminescence spectrum of the same sample measured at the same location. (c) Energy shift of the lowest-energy branch extracted from (a) as a function of $n/n_s$. $n$ is the carrier density and $n_s$ is the full filling density of the first narrow band. (d) The TB calculation of local carrier density difference between the states in the AA and AB/BA regions as a function of $n/n_s$. Inset was a schematic exemplification of the X$_{\mathrm{RM}}$ with the lowest energy confinement on the electron-doped side. (e) TB calculation of the spatial charge distribution of TBG with $\theta=0.6^{\circ}$ at different doping densities. The lowest map was a schematic of relaxed TBG with AA, AB and BA stackings. From Ref. \cite{hu2023observation}. Reprinted with permission from AAAS.}
    \label{fig:exciton}
\end{figure*}

Second, commensurate configurations of TBG with twist angle close to $30^{\circ}$ were used as the approximant. In these approximants, the top graphene layer was compressed or stretched to satisfy the condition $M \times 3d= N \times a_t$, with $a_t$ being the lattice constant of the top graphene with strain. The approximant was named as $M/N$. The structure of 4/7 approximant is shown in Fig. \ref{fig:quasicrystal}(d). The accuracy of these approximants were varified by comparing the DOS and optical conductivity with those calculated directly from the quasicrystal. Moreover, the quasi-periodicity still remained in the periodic approximants. The eigenstates obtained from the approximant perserved the 12-fold rotational symmetry (Fig. \ref{fig:quasicrystal}(e)). The approximant was used to study the magnetic field effect. Some new Landau levels (LLs) appeared below Fermi level by 1.6 eV when the magnetic field exceeded 10 T. These new LLs followed a two-dimensional Dirac fermion with reduced Fermi velocity of 5.21 $\times 10^{5}$ m/s. Moreover, the LL of $n=0$ was missing, but its position was predicted to be around -1.49 eV by interpolation. At this energy, there was a band gap at M point, and the valleys hybridized strongest.    

\subsection{Rydberg moir\'e excitons in WSe\texorpdfstring{$_2$}{2}/TBG heterostructures}
Another example is the observation of the Rydberg moiré excitons in WSe$_2$/TBG heterostructure\cite{hu2023observation,he2024dynamically}. In this system, the induced moir\'e potential in TBG provided a possible pathway to spatially confine and manipulate the Rydberg excitons in the monolayer WSe$_2$. We named the moir\'e-trapped Rydberg excitons as Rydberg moir\'e excitons. For TBG with angle below a crossover angle $\theta=1.2^{\circ}$, the lattice relaxation played a significant role in both structural and electronic properties\cite{gargiulo2017structural,nguyen2021electronic}. In the geometry, the lattice relaxation shrunk the AA region and expanded the AB region to a triangular domain (see the inset of Fig. \ref{fig:exciton}(e)). The states from lowest energy narrow bands were mainly localized in the AA region and states from the remote bands were mainly in the AB region\cite{nguyen2021electronic}. Such lattice reconstruction was relevant in the generation of the Rydberg moir\'e excitons in WSe$_2$/TBG heterostructures. The lattice relaxation effect could be well captured by a combination of molecular dynamics, TB Hamiltonian and the TBPM methods.   

In the WSe$_2$/TBG heterostructure, when the angle in TBG was relatively low, for instance $\theta=0.6^{\circ}$, the period $\lambda$ of the moir\'e pattern was larger than the exciton size $r_B$ ($\approx$ 7 nm for the 2s states in monolayer WSe$_2$\cite{stier2018magnetooptics}). Due to the lattice relaxation, the AA region had a radius of $\approx2.6$ nm (estimated from the half maximum of the spatially accumulated charge peak), much smaller than $r_B$. Moreover, the accumulated charges in the AA region of the TBG were strong enough to trap the opposite charge of the 2s exciton. Then, the system was in a strong coupling regime with $\lambda/r_B > \approx 2.4$. In this regime, the Rydberg moir\'e excitons X$_{\mathrm{RM}}$ showed some significant features in the reflectance spectra (see Fig.\ref{fig:exciton}(a)): 1) multiple energy splittings near 1.783 eV, 2) pronounced red shift, 3) narrowed linewidth, indicating a significant enhancement of the interlayer Rydberg exciton--accumulated charge interactions. Such features were confirmed by photoluminescence measurements in Fig. \ref{fig:exciton}(b). The energy shift magnitude $|E_{shift}|$ from the charge neutrality point (CNP) was extracted, which showed a nonmonotonic dependence on the density. Then, the real-space charge distrubution in TBG was calculated by a combination of the TB Hamiltonian in Eq (\ref{hal0}) with TBPM methods, and molecular dynamics for lattice relaxation\cite{hu2023observation}. As shown in Fig. \ref{fig:exciton}(e), in the CNP, the local charge density located mainly in the AA region, which created deep and narrow potential wells for trapping charges of the exciton. The $|E_{shift}|\approx(eU_{AA}-eU_{AB/BA})\propto(n_{AA}-n_{AB/BA})$ estimated from the difference in attraction in the AA region and repulsion in the AB/BA region, is plotted in Fig. \ref{fig:exciton}(d). The nonmonotonic trend was similar to the observed result. 


\section{Summary and perspectives}
We have carefully reviewed the single-particle, atomistic TB Hamiltonian for twisted graphene layers. Intralayer and interlayer hoppings in graphene-based moiré materials can be described by the Slater--Koster relation. The single-particle TB Hamiltonian can be combined with Hartree--Fock interactions and a Hubbard-$U$ term within a mean-field approximation. A rescaling strategy can reduce the computational cost of self-consistent mean-field calculations. The SK relation including the $p_z$ orbital remains valid when constructing TB Hamiltonians for hBN-based moiré materials, though the hopping parameters fitted from DFT differ from those of graphene-based systems. For TMD-based moiré materials, an \textit{ab initial} intralayer TB Hamiltonian is needed, while SK relations can be employed to generate an interlayer Hamiltonian. Beyond traditional diagonalization methods, robust linear-scaling approaches can be combined with real-space atomistic TB Hamiltonians to compute diverse properties of moiré materials. Machine-learning methods are accelerating the construction of \textit{ab initial}-quality TB Hamiltonians for moiré systems. We also summarized how low-energy continuum models can be derived from atomistic TB models. Other low-energy effective lattice models are crucial for understanding electron--electron interaction phenomena in moiré superlattices, but lie beyond the scope of this work.\cite{koshino2018maximally,kang2018symmetry,po2018origin,yuan2018model,dodaro2018phases,wu2018hubbard}

As for future prospects of atomistic TB methods for simulating moir\'e materials, an essential direction is the accurate parameterization of TB Hamiltonians for systems not only with hexagonal lattices (the main focus here) but also with rectangular, kagome, and more general lattices\cite{yu2025general}, and searching for moiré flat bands in other 2D superlattices. As more experimental results of correlated phases and topology are reported, the TB method is still an accurate enough and powerful tool to understand the origin of the flat band-related correlated phenomena, and needs to be further explored. Building open databases for training deep-learning Hamiltonian models\cite{bao2024deep} will further facilitate data-driven construction and discovery of new interesting moir\'e superlattices. From the perspective of practice, for simulations of large-scale moir\'e systems, linear-scaling random-state methods require additional development to ensure compatibility with TB Hamiltonians in non-orthogonal basis.

\section*{Conflicts of interest}
There are no conflicts to declare.

\section*{Data availability}
No primary research results, software or code have been included and no new data were generated or analysed as part of this review.

\section*{Acknowledgements}
IMDEA Nanociencia acknowledges support from the ‘Severo Ochoa’ Programme for Centres of Excellence in R\&D (CEX2020-001039-S/AEI/10.13039/501100011033). PAP, FG and ZZ acknowledge support from NOVMOMAT, project PID2022-142162NB-I00 funded by MICIU/AEI/10.13039/501100011033 and by FEDER, UE as well as financial support through the (MAD2D-CM)-MRR MATERIALES AVANZADOS-IMDEA-NC. ZZ acknowledges support from the European Union's Horizon 2020 research and innovation programme under the Marie-Sklodowska Curie grant agreement No 101034431. FE acknowledges support funding from the European Union's Horizon 2020 research and innovation programme under the Marie Skłodowska-Curie grant agreement No 101210351. PAP acknowledges funding by Grant No.\ JSF-24-05-0002 of the Julian Schwinger Foundation for Physics Research.



\balance


\bibliography{rsc} 
\bibliographystyle{rsc} 

\end{document}